\documentclass[aps,prx,twocolumn,showpacs,notitlepage,superscriptaddress,letterpaper]{revtex4-2}
\usepackage{graphicx, amsmath, amssymb, amsfonts, pifont, bm, cancel}
\usepackage[usenames]{color}
\usepackage{subfigure}
\usepackage[normalem]{ulem} 
\usepackage{circuitikz}
\usepackage{hyperref}
\usepackage{upgreek}
\usepackage{outlines}
\usepackage{enumitem}
\usepackage{blkarray}

\setenumerate[1]{label=\Roman*.}
\setenumerate[2]{label=\Alph*.}
\setenumerate[3]{label=\roman*.}
\setenumerate[4]{label=\alph*.}

\definecolor{MyDarkBlue}{rgb}{0,0,1}
\definecolor{darkgreen}{rgb}{0.0, 0.5, 0.0}
\usepackage{physics}

\begin{document}

\title{A Tunable, Modeless, and Hybridization-free Cross-Kerr Coupler for Miniaturized Superconducting Qubits}

\author{Gihwan Kim}
\affiliation{Kavli Nanoscience Institute and Thomas J. Watson, Sr., Laboratory of Applied Physics, California Institute of Technology, Pasadena, California 91125, USA.}
\affiliation{Institute for Quantum Information and Matter, California Institute of Technology, Pasadena, California 91125, USA.}

\author{Andreas Butler}
\affiliation{Kavli Nanoscience Institute and Thomas J. Watson, Sr., Laboratory of Applied Physics, California Institute of Technology, Pasadena, California 91125, USA.}
\affiliation{Institute for Quantum Information and Matter, California Institute of Technology, Pasadena, California 91125, USA.}

\author{Oskar~Painter}
\email{opainter@caltech.edu}
\homepage{http://painterlab.caltech.edu}
\affiliation{Kavli Nanoscience Institute and Thomas J. Watson, Sr., Laboratory of Applied Physics, California Institute of Technology, Pasadena, California 91125, USA.}
\affiliation{Institute for Quantum Information and Matter, California Institute of Technology, Pasadena, California 91125, USA.}

\date{\today}
\begin{abstract}
Superconducting quantum circuits typically use capacitive charge-based linear coupling schemes to control interactions between elements such as qubits. While simple and effective, this coupling scheme makes it difficult to satisfy competing circuit design requirements such as maintaining large qubit anharmonicity and coherence along with a high degree of qubit connectivity and packing density. Moreover, tunable interactions using linear coupling elements produce dynamical variations in mode hybridization, which can induce non-adiabatic transitions, resulting in leakage errors and limiting gate speeds. In this work we attempt to address these challenges by proposing a junction-based coupling architecture based on SQUID (superconducting quantum interference device) couplers with relatively small Josephson energies. SQUID couplers provide intrinsic cross-Kerr interactions that can be controlled by external fluxes and that do not rely on mode hybridization. The small Josephson energies of the coupler maintain the interaction at a perturbative scale, which limits undesired higher-order mixing between coupled elements while achieving a sufficiently strong cross-Kerr interaction originating from diagonal coupling elements. Based on these properties, we show that a SQUID coupler can be used to implement a fast, adiabatic, and high-fidelity controlled-Z gate without introducing extra modes, and the operation is robust against junction asymmetry for high-frequency qubits. Although unconventional crosstalk may arise due to junction asymmetries and parasitic hybridization with spectator qubits, we show that these effects are sufficiently small for realistic circuit parameters. As an example of the utility of such junction-based coupling schemes, we present a scalable tiling strategy for a miniaturized superconducting quantum processor based on merged-element transmon qubits.

\end{abstract}
\maketitle

\clearpage


\section{Introduction}
\label{Section:Introduction}

Superconducting quantum processors utilize capacitors to realize linear couplings between circuit elements such as qubits, tunable couplers, and readout resonators \cite{koch2007firsttransmon, yan2018tunable, stehlik2021ibmcoupler,campbell2023modular, goto2022doubletransmon, krantz2019aquantum}. However, the design of capacitances can be constrained due to other goals, such as suppressing dielectric loss \cite{martinis2005decoherence, lisenfeld2019electric, murray2021materialmatters}, maintaining large qubit anharmonicity \cite{koch2007firsttransmon, krantz2019aquantum}, and optimizing the form-factor and size of qubits~\cite{gidney2025factor2048bitrsa}. Furthermore, tunable coupling schemes that utilize linear coupling elements rely on temporal variations in mode hybridization, which can introduce leakage errors through non-adiabatic transitions, thereby limiting gate speed and fidelity \cite{martinis2014fastadiabatic, yan2018tunable,sung2021zzfreeiswap}.

Merged-element transmons (mergemons) \cite{zhao2020mergemon, mamin2021mergemondesign, daum2025investigationparasitictwolevelsystems} provide a representative example of the challenges associated with using linear capacitive coupling. The concept behind the mergemon qubit is to localize the qubit capacitance to the small volume of the Josephson junction, limiting the (spectral) density of two-level system defects (TLSs) \cite{martinis2005decoherence, lisenfeld2019electric, murray2021materialmatters} that can decohere the qubit. Additional geometric capacitance in the circuit, for example to implement coupling between qubits and coupling to resonators for qubit read-out, increases the number of interacting TLS. Reducing the strength of qubit-TLS interactions can be accomplished by enlarging the capacitor volume and diluting the electric field intensity, however this leads to larger and less confined qubits that are more prone to parasitic crosstalk with other circuit elements~\cite{blais2004circuitqed, koch2007firsttransmon}.



To overcome the bulkiness and mode hybridization of standard capacitive coupling architectures, we propose using SQUID (superconducting quantum interference device) couplers for dynamic coupling of elements. This approach is similar to other recent junction-based coupler proposals~\cite{kounalakis2018tunable, ye2021quarton, chapple2025balanced, wang2025longitudinal, stolyarov2025twophoton, goto2022doubletransmon, campbell2023modular, li2024doubletransmon, chakraborty2025tunablesuperconductingquantuminterference, brooks2013protected, hays2025nondegeneratenoiseresilientsuperconductingqubit}, with the distinct difference being that we focus on SQUIDs composed of Josephson junctions with relatively small tunneling energies to preserve a desired hierarchy of interactions and to limit the effects of junction non-idealities and noise. A primary benefit of a junction-based coupling scheme is that it need not introduce any additional modes to the system, modes that can serve as a channel for leakage errors when implementing qubit gate operations~\cite{mcewen2021removing, miao2023overcoming}. In conventional approaches using linear coupling elements, entangling interactions arise indirectly from nonlinearities of the qubits themselves~\cite{chen2014gmon, yan2018tunable, sung2021zzfreeiswap, stehlik2021ibmcoupler, goto2022doubletransmon, li2024doubletransmon}. In contrast, the SQUID coupler directly introduces intrinsic cross-Kerr interactions that can be switched on or off using external magnetic flux. As the cross-Kerr interaction does not rely on hybridization of the interacting elements, mixing between qubits can be independently engineered and suppressed \cite{chapple2025balanced}. In addition, the removal of geometric coupling capacitors relaxes the trade-off between dielectric loss and qubit size.

Along with the potential benefits of the SQUID coupler, there are several technical challenges that must also be considered. In particular, the effect of fabrication-induced junction asymmetries and the presence of extraneous SQUID loops formed via ground connections of the coupler can limit the coupler performance~\cite{braumuller2020squidgeometryperimeter,hays2025nondegeneratenoiseresilientsuperconductingqubit}. Junction asymmetry in the SQUID coupler can lead to a reduced on-off extinction ratio, crosstalk with spectator qubits, and in conjunction with the extraneous SQUID loops, an increased sensitivity to magnetic flux noise. SQUID couplers used with floating transmons eliminate the presence of extraneous outer SQUID loops~\cite{kounalakis2018tunable}; however, such a design introduces sloshing modes that can mediate parasitic interaction and leakage errors. In this work we take a different approach, and show that SQUID couplers formed from junctions with relatively small Josephson tunneling frequencies ($\lesssim 1$~GHz) compared to the transition frequencies of the elements being coupled together ($\sim 5$~GHz), and for junction asymmetries up to $20\%$, can achieve cross-Kerr interaction rates of a few tens of MHz with high extinction ratio and low parasitic crosstalk to neighboring elements, while maintaining low infidelity due to typical flux noise levels. 


An outline of our analysis of the proposed SQUID coupler is as follows. We begin in Sec.~\ref{Section:CouplerTwoQubitGate} with a review of the Hamiltonian and external flux control parameters for a SQUID coupler mediating coupling between two detuned transmon qubits, and show through numerical simulation of a quantized circuit model of a system with relatively small coupler junction energies, that a high-fidelity and fast controlled-Z (CZ) gate can be implemented with minimal adiabaticity overhead. In Sec.~\ref{Section:Robustness} we study both perturbatively and numerically the robustness of the CZ gate performance to coupler junction asymmetry. We show that coherent errors remain below $5 \times 10^{-7}$ for gate times of $22$~ns over the entire range of junction asymmetry.  A sensitivity analysis to flux noise is then presented in Sec.~\ref{Section:Sensitivity}, indicating that for small Josephson energies and relatively high qubit frequencies that the entangling phase of the CZ gate is robust to typical levels of flux noise in superconducting circuits. In Sec.~\ref{Section:Crosstalk} we identify two novel forms of crosstalk that can be mediated by the SQUID couplers, and show that both forms of crosstalk remain sufficiently small for realistic circuit parameters. Finally, in Sec.~\ref{Section:Tiling} we present a scalable mergemon architecture where all interactions are mediated by junction-based SQUID couplers, highlighting the potential for realizing a fully miniaturized quantum processor. 

\section{Tunable cross-Kerr Interaction via SQUID coupler}
\label{Section:CouplerTwoQubitGate}

In order to elucidate the core aspects of the proposed coupling scheme we begin with an analysis of the circuit illustrated in Fig.~\ref{fig:SQUIDCoupler}a, comprising two transmon qubits coupled by a SQUID coupler. $E_{J,i}$, $E_{C, i}$, and $C_{i}$ are the Josephson energy, the charging energy, and the capacitance to ground of the transmon $i \in \{1, 2\}$, respectively,  $E_{J,C1}, E_{J,C2} \ll E_{J,1}, E_{J, 2}$ and $C_{C1}, C_{C2} \ll C_1, C_2$ are the Josephson energies and the capacitances of the junctions in the SQUID coupler. $\Phi_{e,1}$ and $\Phi_{e,2}$ are external fluxes that thread the inner SQUID loop within the SQUID coupler and the outer SQUID loop formed along the ground connection, respectively. 


The Hamiltonian describing the circuit is given by:

{
\begin{align}
    \hat{H}_{\text{tot}} &=  \hat{H}_1 + \hat{H}_2+ \hat{H}_{\text{int}}, \label{eq:Hamiltonian_start}\\
    \hat{H}_i &=  4E_{C,i}\hat{n}_i^2 - E_{J,i}\cos{\hat\varphi_i} \  \left(i \in \{1,2\}\right), \label{eq:Hamiltonian_i} \\ 
    \hat{H}_\text{int} &= \notag  -E_{J,C1}\cos{\left(\hat\varphi_2 - \hat\varphi_1 + \varphi_{e,1}+ \varphi_{e,2}\right)} \notag \\-&E_{J,C2}\cos{\left(\hat\varphi_2 - \hat\varphi_1 + \varphi_{e,2}\right)} + g \hat{n}_1\hat{n_2},
\label{eq:Hamiltonian_end}
\end{align}
}

\noindent where $\hat{\varphi}_{1}$, $\hat{\varphi}_{2}$, $\hat{n}_{1}$, and $\hat{n}_{2}$ are phase and charge operators, $g = 4e^2C_C/C^2$ denotes the charge coupling rate due to the capacitances of the coupler junctions where $C_C \equiv C_{C1}+C_{C2}$, $C^2 \equiv C_1C_2 + C_CC_1 + C_CC_2$, and $\varphi_{e,k} \equiv 2\pi \Phi_{e, k}/\Phi_0 \ (k \in \{1,2\})$ are the reduced external fluxes. Details of the circuit quantization are provided in Appendix \ref{App:circuitquantization}. 

We introduce the flux ``operating condition", $\varphi_{e,1} + 2\varphi_{e,2}=0$, which ensures that the even-parity interactions, such as excitation hopping and cross-Kerr interaction, dominate over odd-parity interactions, such as longitudinal and two-photon exchange interactions. Under this constraint, the interaction Hamiltonian can be rewritten using trigonometric relations as follows:

\begin{figure}[tbp]
\centering
\includegraphics[width = \columnwidth]{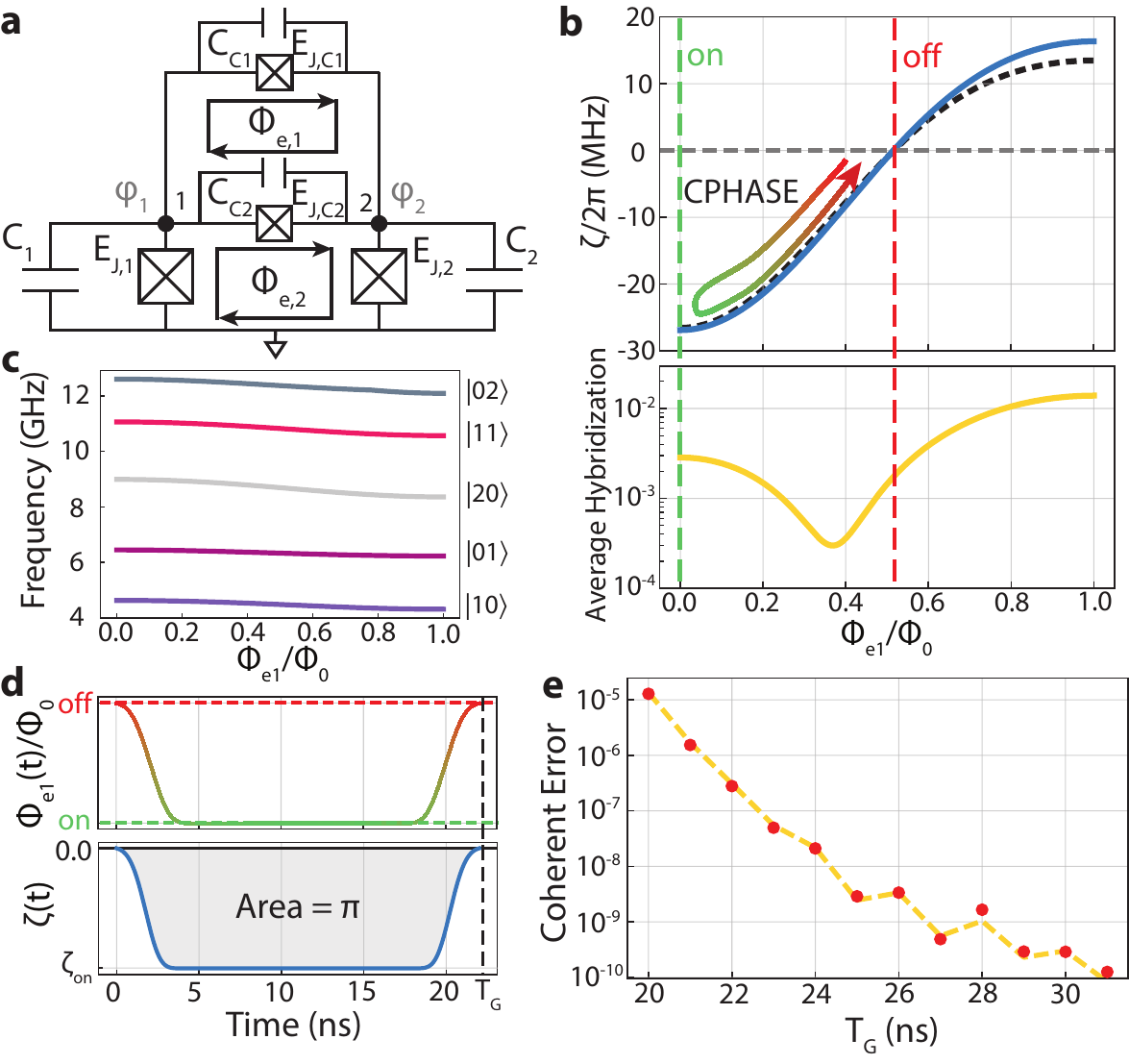}
\caption{\textbf{Tunable cross-Kerr coupling and CZ gate via a SQUID coupler.} \textbf{a}, Circuit schematic of two transmons coupled via a SQUID coupler. \textbf{b}, ZZ interaction rate $\zeta$ and average hybridization as functions of $\Phi_{e,1}$. Red (green) dashed line indicates where $\zeta$ is zero (maximal). Black dashed line represents $\zeta$ calculated from perturbation theory. \textbf{c}, Eigenfrequencies of the first 5 excited states as functions of $\Phi_{e,1}$ under the operating condition $\Phi_{e,2} = -\Phi_{e,1}/2$. \textbf{d}, Flux waveform $\Phi_{e,1}(t)$ and the corresponding $\zeta(t)$, used for a 22 ns-long CZ gate. \textbf{e}, Coherent error of CZ gate for various gate durations $T_\text{G}$ (red circles). Yellow dashed line represents the contribution from non-adiabatic state transitions.}
\label{fig:SQUIDCoupler}
\end{figure}

\begin{align}
    &\hat{H}_\text{int}|_{\varphi_{e,1} + 2\varphi_{e,2} = 0} = -\Sigma E_{J,C}\cos{\left(\frac{\varphi_{e,1}}{2}\right)}\cos{(\hat{\varphi}_2-\hat{\varphi}_1)}\notag \\ & \ \ \ \ \ \ +\Delta E_{J,C}\sin{\left(\frac{\varphi_{e,1}}{2}\right)}\sin{(\hat{\varphi}_2-\hat{\varphi}_1)} + g \hat{n}_1\hat{n}_2,
    \label{eq:Hint_flux_removed}
\end{align}

\noindent where $\Sigma E_{J,C} \equiv E_{J,C1} + E_{J,C2}$ and $\Delta E_{J,C} \equiv E_{J,C1} - E_{J,C2}$ are the total and differential Josephson energies. The cosine term proportional to $\Sigma E_{J,C}$ provides even-parity interactions, whereas the sine term with $\Delta E_{J,C} \ll \Sigma E_{J,C}$ provides odd-parity interactions. We first focus on symmetric SQUID couplers ($\Delta E_{J,C} = 0$), which leave only the even-parity terms. The effects of junction asymmetry are discussed in the following sections.

As $E_{J,C}$s are small, perturbation theory can be used to calculate the effective linear coupling rate, $g_\text{eff} \equiv \langle10|\hat{H}_\text{int}|01\rangle$, and the first- and second- order corrections, $\zeta^{(1)}$ and $\zeta^{(2)}_{c}$, from the excitation-number-conserving matrix elements to the ZZ interaction rate, $\zeta \equiv \omega_{\tilde{|00\rangle}}-\omega_{\tilde{|01\rangle}}-\omega_{\tilde{|10\rangle}}+ \omega_{\tilde{|11\rangle}}$:

\begin{align}
    g_\text{eff} &= -\Sigma E_{J,C}'\cos{\left(\frac{\varphi_{e,1}}{2}\right)}\varphi_1^{zpf}\varphi_2^{zpf} + gn_1^{zpf}n_2^{zpf}, \label{eq:effectivecoupling}\\
    \zeta^{(1)} &= -\Sigma E_{J,C}'\cos{\left(\frac{\varphi_{e,1}}{2}\right)} (\varphi_1^{zpf})^2(\varphi_2^{zpf})^2,
    \label{eq:firstorder} \\
    \zeta^{(2)}_{c} &= \frac{|\langle02|H_\text{int}|11\rangle|^2}{\omega_{|11\rangle} - \omega_{|02\rangle}} +  \frac{|\langle20|H_\text{int}|11\rangle|^2}{\omega_{|11\rangle} - \omega_{|20\rangle}} \approx \frac{4g_\text{eff}^2\eta }{\Delta^2 - \eta^2}. 
    \label{eq:perturbation}
\end{align}

\noindent Here $\Sigma E_{J,C}' \equiv \Sigma E_{J,C}e^{-\frac{(\varphi_1^{zpf})^2}{2}}e^{-\frac{(\varphi_2^{zpf})^2}{2}}$ captures the byproduct of normal ordering, $\eta_1, \eta_2 \approx \eta$ are bare transmon anharmonicities, $\omega_{1}, \omega_{2}$ are bare transmon ground-to-first-excited state transition frequencies, and $\Delta \equiv \omega_{1} - \omega_{2}$ is the transmon qubit detuning. $|ij\rangle$, $i, j \in \{0,1\}$ represents the bare eigenstate, where the first (second) index refers to transmon 1 (transmon 2) excitation number. Bare eigenstates and eigenfrequencies are obtained from $\hat{H}_1$ and $\hat{H}_2$. $\omega_{\tilde{|ij\rangle}}$ are the corresponding dressed eigenstate and eigenfrequency. Cosinusoidal terms are expanded in terms of creation ($\hat{a}_{i}^\dagger$) and annihilation operators ($\hat{a}_{i}$), with corresponding phase and charge operators $\hat\varphi_i = \varphi^{zpf}_i(\hat{a}_i + \hat{a}_i^\dagger)$ and $\hat{n}_i = in^{zpf}_i (\hat{a}_i - \hat{a}_i^\dagger)$, where $\varphi_{i}^{zpf}$ and $n_{i}^{zpf}$ are zero-point fluctuations. See Appendix \ref{App:perturbation} for further details.

The first-order correction $\zeta^{(1)}$ originates from diagonal perturbation terms, reaching a maximum at $\Phi_{e,1} = 0$ (``on") and zero at $\Phi_{e,1} = 0.5\Phi_0$ (``off"). Unlike ZZ interactions arising from hybridization, $\zeta^{(1)}$ depends only weakly on the detuning, allowing strong ZZ interaction rates even with a large detuning. The effective linear coupling $g_\text{eff}$ can be reduced and can even be canceled out in the range $\Phi_{e,1} \in [0.0\Phi_0, 0.5\Phi_0)$ with appropriate choices of $C_C$ due to the relative signs of the terms in eq.~(\ref{eq:effectivecoupling}) \cite{kounalakis2018tunable, campbell2023modular, chapple2025balanced}. With sufficiently small $g_\text{eff}$ and large detuning, hybridization-induced ZZ interactions such as $\zeta^{(2)}_{c}$ can be suppressed in a substantial range of $\Phi_{e,1}$. Under such conditions, $\zeta^{(1)}$ dominates the ZZ interaction rate, and there is at least one idle external flux $\Phi_\text{off}$ at which the ZZ interaction can be turned off $\zeta(\Phi_\text{off}) = 0$ and the two transmons can idle. 

\begin{table}[tbp]
\centering
\renewcommand{\arraystretch}{1.3}
\begin{tabular}{|c | c| c| c|}
\hline
$E_{J,1}/2\pi$ & 11.5 GHz & $E_{J,2}/2\pi$ & 20.0 GHz\\
$C_{1}$ & 77.5 fF& $C_{2}$ & 69.2 fF \\
$E_{J,C1}/2\pi$  & 0.40 GHz & $E_{J,C2}/2\pi$  & 0.40 GHz\\
$C_{C1}$ & 0.78 fF & $C_{C2}$ & 0.78 fF \\
\hline
$\omega_{1}/2\pi$ & 4.49 GHz & $\omega_{2}/2\pi$ & 6.33 GHz  \\
$\eta_{1}/2\pi$ & -0.284 GHz & $\eta_{2}/2\pi$ & -0.306 GHz\\
$\Phi_\text{off}$  & 0.516 $\Phi_0$ & $\zeta_\text{on}/2\pi$ & -26.9 MHz
\\
\hline
\end{tabular}
\caption{Parameters used for numerical analysis. Parameters indicating energies are normalized by the reduced Planck constant $\hbar$.}
\label{Table:circuitparams}
\end{table}

In Fig.~\ref{fig:SQUIDCoupler} we present a specific numerical example of the tunable cross-Kerr coupling between two far-detuned transmon qubits for a set of coupler and qubit parameters listed in Table~\ref{Table:circuitparams}. Note that junction tunneling energies and capacitances in this table are chosen to be consistent with experimentally achievable junction parameters~\cite{josephson1962possible, stewart1968currentvoltage, mccumber1968effectjosephson, zhao2020mergemon, mamin2021mergemondesign, chen2024phonon, daum2025investigationparasitictwolevelsystems,mamin2021mergemondesign, daum2025investigationparasitictwolevelsystems}. Here we directly simulate the full Hamiltonian in eq.~(\ref{eq:Hamiltonian_start}) by quantizing the circuit model and without applying rotating-wave approximations~\cite{Groszkowski2021scqubitspython, johansson2012qutip, allen2012optical}.
Fig.~\ref{fig:SQUIDCoupler}b shows $\zeta$ as a function of $\Phi_{e,1}$. An idle external flux $\Phi_\text{off} = 0.516\Phi_0$ and a large ZZ interaction rate $\zeta_\text{on}/2\pi = -26.9$ MHz at $\Phi_\text{on} = 0$ are found. Although $\zeta$ varies appreciably, the change in eigenfrequencies remains less than 320 MHz throughout the tuning range for both qubits, as shown in Fig.~\ref{fig:SQUIDCoupler}c, minimizing susceptibility to flux noise \cite{koch2007firsttransmon, bylander2011noise,krantz2019aquantum, garcia2022weaklytunable}. We chose capacitances $C_{C}/2 = C_{C1} = C_{C2}$ of the coupler that minimizes hybridization with other states in the relevant tuning range $[0, \Phi_\text{off}]$. To quantify this, the average hybridization with other states $\sum_{mn} (1-P_{\ket{mn}})/4$ is used, where $P_{\ket{mn}} = |\langle mn | \tilde{mn}\rangle|^2$ is the overlap between bare ($|{mn}\rangle$) and dressed ($|\tilde{mn}\rangle$) eigenstates. As shown in Fig.~\ref{fig:SQUIDCoupler}b, the average hybridization is below 0.3\% throughout the range $[0, \Phi_\text{off}]$. 

A controlled phase (CPHASE) can be implemented by dynamically tuning the external flux from the idle bias point $\Phi_\text{off}$ to $\Phi_\text{on}$, and back to $\Phi_\text{off}$, as illustrated in Fig.~\ref{fig:SQUIDCoupler}d. To implement a CZ gate we aim for the accumulated controlled phase to be equal to $\pi$, i.e., $-\int_0^{T_G}\zeta(t) dt = \pi \  (\text{mod} \  2\pi)$, where $T_G$ is the total gate duration and $\zeta(t)$ is the instantaneous ZZ interaction rate. We use a simplified adiabatic control technique similar to that employed in Refs.~\cite{martinis2014fastadiabatic, baksic2016speedingupadiabatic} by treating $\ket{11}$ and $\ket{02}$ as a two-level system, in which the rate of change in hybridization (mixing-angle) between the two states is tailored. A target mixing-angle waveform is first generated by convolving a square pulse with a duration of $\beta T_G$ and a Slepian-like pulse with a duration of $(1-\beta)T_G$, in which $\beta$ controls the adiabaticity while keeping $T_G$ fixed. The flux waveform is then obtained by mapping the mixing-angles to external flux, and subsequently low-pass filtered by convolving with a Gaussian kernel of $\sigma = 0.5$ ns to account for realistic flux signal conditioning~\cite{lacroix2023fast, hellings2025calibratingmagneticfluxcontrol}. An example waveform $\Phi_{e,1}(t)$ and the corresponding $\zeta(t)$ are shown in Fig.~\ref{fig:SQUIDCoupler}d. 

The coherent error of the CZ gate as a function of $T_G$ is shown in Fig.~\ref{fig:SQUIDCoupler}e. For each $T_G$, $\beta$ is calibrated solely to achieve the target entangling phase. Under such calibration, non-adiabatic state transitions between states that are relatively close in frequency, such as $|\tilde{10}\rangle \leftrightarrow |\tilde{01}\rangle$ and $|\tilde{11}\rangle \xrightarrow{} |\tilde{20}\rangle, |\tilde{02}\rangle$, are dominant sources of errors, which can be reduced by increasing $T_G$. We find that it is possible to implement a CZ gate with coherent errors below $3\times10^{-7}$ in 22 ns, showing a minimal adiabaticity overhead of approximately 3.4 ns compared to the minimum gate time $\text{min} \ T_G=\pi /|\zeta_\text{on}| \approx 18.6$ ns. This overhead may be further reduced through more advanced adiabatic control techniques~\cite{baksic2016speedingupadiabatic} and less strict filtering. Note that the simulation uses the full Hamiltonian without employing the rotating-wave approximation. The gate is also simulated in the presence of $T_1$ relaxation errors, where a coherence-limited infidelity $1-F$ of approximately $1.8\times 10^{-5} \approx 0.8T_G/T_1$ is found when assuming $T_1 = 1 \ \text{ms}$ for both transmons \cite{pedersen2007fidelity} (see Appendix~\ref{App:gatefidelityT1}). 


\section{Robustness against junction asymmetry}
\label{Section:Robustness}

The odd-parity interaction is proportional to $\Delta E_{J,C}\sin{(\varphi_{e,1}/2)}$, reaching its strongest value near the idle point of $\Phi_\text{off} \approx 0.5\Phi_0$. This term includes longitudinal $(\hat{a}^\dagger_i\hat{a}_i (\hat{a}^\dagger_j +\hat{a}_j))$ and two-photon exchange $((\hat{a}^\dagger_i)^2 \hat{a}_j + h.c.)$ interactions. The contribution to the ZZ interaction rate from the odd-parity interaction, $\zeta_\text{odd}^{(2)}$, scales inversely with odd-parity combinations of qubit frequencies and proportionally to higher-orders of phase zero-point fluctuations, 

\begin{equation}
    \zeta_\text{odd}^{(2)} \sim \mathcal{O}\left(\frac{\Delta E_{J,C}^2(\varphi^{zpf})^{6}}{m\omega_1 + n\omega_2}\right) + \mathcal{O}\left(\frac{\Delta E_{J,C}^2(\varphi^{zpf})^{4}}{\omega^2}\eta\right)\notag,
\end{equation}

\noindent where $(m, n) \in \{(1, 0), (0, 1), (2, -1), (2, -1)\}$. A detailed calculation is provided in Appendix \ref{App:perturbation}. Note that $(\varphi^{zpf})^6 \approx (2E_C/E_J)^{3/2}$ is less than $0.02$ for a typical transmon where $E_J/E_C > 30$. Thus, for a SQUID coupler with junction tunneling energies ($E_{J,C}$) much less than the qubit frequencies, and for transmon qubits with relatively small $\varphi^{zpf}$, we expect that contributions to the ZZ interaction from the odd-parity interaction will be strongly suppressed, preserving the existence of an idle point, $\Phi_\text{off}$. 

\begin{figure}[tbp]
\centering
\includegraphics[width = \columnwidth]{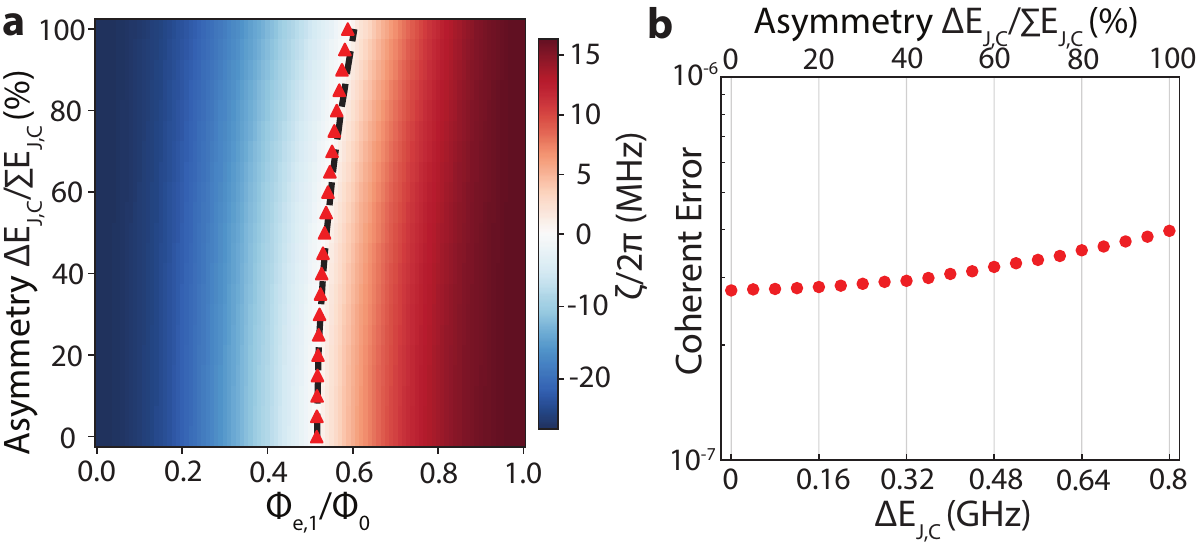}
\caption{\textbf{Robustness against junction asymmetry.} \textbf{a}, ZZ interaction rate as a function of $\Phi_{e,1}$ and junction asymmetry. Junction capacitance asymmetry is set to be the same as the Josephson energy asymmetry. Red triangles indicate $\Phi_\text{off}$ obtained from circuit quantization, and black dashed line shows $\Phi_\text{off}$ estimated from perturbation theory. \textbf{b}, Coherent error of the 22 ns-long CZ gate as a function of junction asymmetry. 
}
\label{fig:asymmetry}
\end{figure}

In Fig.~\ref{fig:asymmetry} we study the effects of junction asymmetry in the SQUID coupler for the nominal circuit parameters of Table~\ref{Table:circuitparams}. Using circuit quantization, $\zeta$ is estimated as a function of junction asymmetry $\Delta E_{J,C}/\Sigma E_{J,C}$ while $\Sigma E_{J,C}$ is kept constant, as shown in Fig.~\ref{fig:asymmetry}a. Josephson energy asymmetries and junction capacitance asymmetries are assumed to be the same. As expected from the analysis above, we find that the ZZ interaction rate does not depend significantly on junction asymmetry, and an idle external flux exists over the entire range of $\Delta E_{J,C}$. Estimates of $\Phi_\text{off}$ from first- and second-order perturbation theory considering up to a single excitation difference in the total excitation number (black dashed line) shows good correspondence with the full numerical computation from circuit quantization (red triangles). The effect of junction asymmetry is further tested by estimating the effect on the CZ gate coherent error, as shown in Fig.~\ref{fig:asymmetry}b. For the 22 ns-long CZ gate, coherent errors remain below $5\times10^{-7}$ for the entire range of possible junction asymmetry (100\% asymmetry corresponding to a single junction coupler as in Ref.~\cite{campbell2023modular, goto2022doubletransmon}). See Appendix \ref{App:circuitparameters} for further details. 

\section{Sensitivity to Flux Noise}
\label{Section:Sensitivity}

\begin{figure}[tbp]
\centering
\includegraphics[width = \columnwidth]{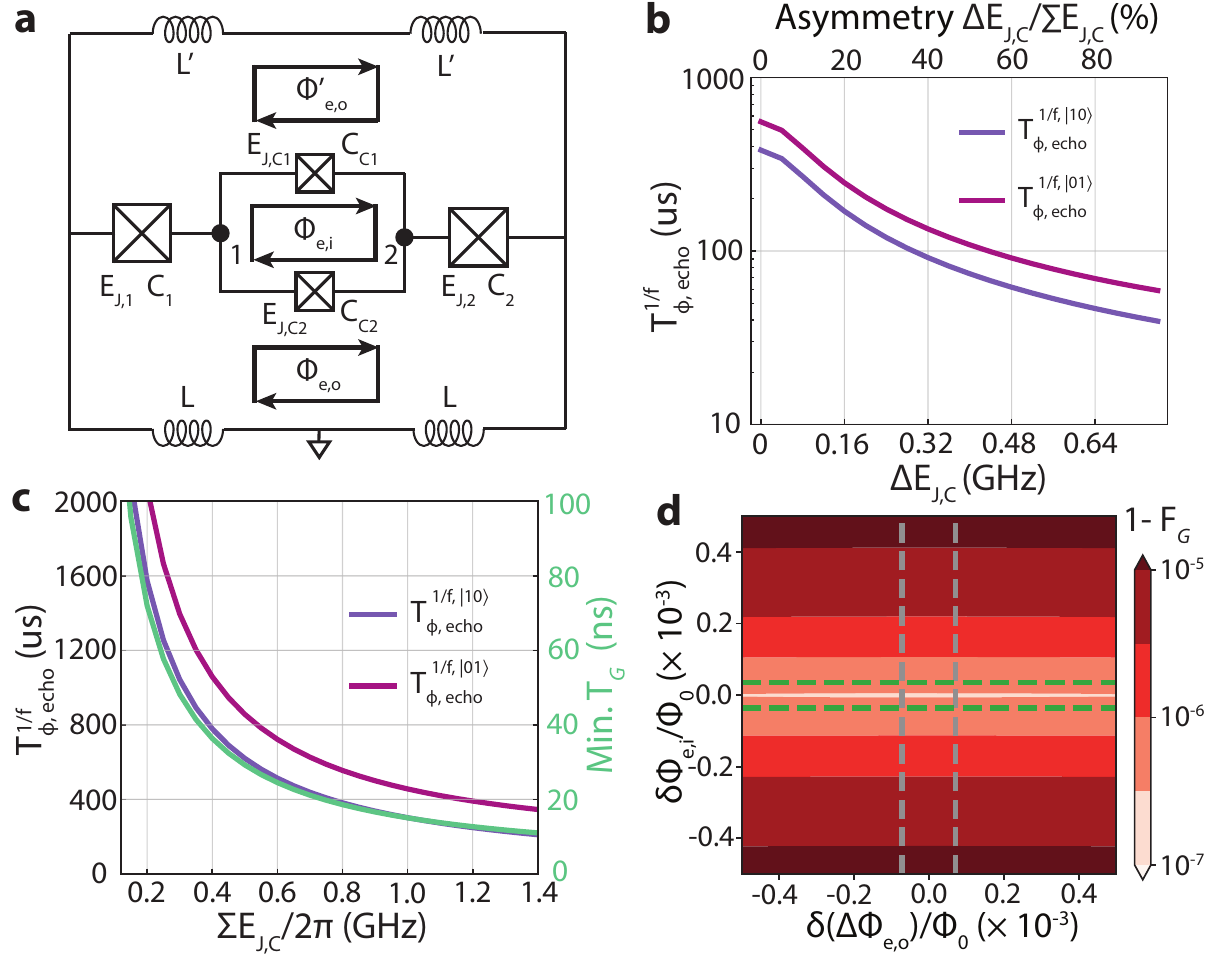}
\caption{\textbf{Sensitivity to flux noise.} \textbf{a}, SQUID coupler circuit with physical external fluxes and realistic inductive network. Capacitors are omitted for brevity. \textbf{b}, Echo dephasing times  $T_{\phi, \text{echo}}^{1/f, \ket{10}}$ and $T_{\phi, \text{echo}}^{1/f, \ket{01}}$ calculated for $1/f$ flux noise with amplitudes $A_{\Phi_{e,i}}=10^{-6}\Phi_0$ and $A_{\Phi_{e,o}}=A_{\Phi_{e,o}'}=5\times10^{-6}\Phi_0$. Here we assume worst-case scenario in which the outer SQUID flux noises are anti-correlated. \textbf{c}, Calculated echo dephasing time and minimum achievable gate time ($\pi/\zeta_\text{on}$) as a function of $\Sigma{E_{J,C}}$ for $1/f$ flux noise with the same flux noise amplitudes as in \textbf{b}. \textbf{d}, Coherent error of the 22 ns-long CZ gate as a function of external flux offsets, $\delta\Phi_{e,1}$ and $\delta\Phi_{e,2}$. Green and gray dashed lines represent the range of flux offsets corresponding to the RMS deviation over a 1 hr drift period for $1/f$ flux noise with amplitudes $A_{\Phi_{e,i}}=A_{\Phi_{e,o}}=A_{\Phi_{e,o}'}=5\times 10^{-6}\Phi_0$, assuming an anti-correlation in the outer SQUID flux noises.
}
\label{fig:sensitivity}
\end{figure}

In order to accurately assess the sensitivity of the proposed junction coupler architecture to magnetic flux noise, the mapping from physical external fluxes to circuit flux variables $\Phi_{e,1}$ and $\Phi_{e,2}$ must account for the inductance in galvanic ground connections~\cite{Schwarz2013gradiometricfluxqubit,braumuller2016concentric,bright2025gradiometric}. This is particularly important when considering the tiling of qubits and SQUID couplers on a two-dimensional lattice (see Sec.~\ref{Section:Tiling}), where the multiple pathways to ground lead to additional extraneous SQUID loops. For such an analysis we therefore consider the SQUID coupler circuit illustrated in Fig.~\ref{fig:sensitivity}a with inductances $L$ and $L'$ much smaller than Josephson inductances of the junctions, modeling realistic ground connections in a square lattice tiling. Accounting for these ground connections we see that the resultant circuit has an inner SQUID loop threaded by external flux $\Phi_{e,i}$ and two outer SQUID loops threaded by external fluxes $\Phi_{e,o}$ and $\Phi_{e,o}'$, with associated reduced external flux variables $\varphi_{e,i}$, $\varphi_{e,o}$, and $\varphi_{e,o}'$. Refer to Appendix \ref{App:circuitquantization} for more details. 

At $L=L'$, the circuit Hamiltonian depends only on $\Phi_{e,i}$ and the differential outer external flux $\Delta\Phi_{e,o}\equiv \Phi_{e,o} - \Phi_{e,o}'$, resulting in what is called a ``gradiometric SQUID coupler". The mapping between the flux parameterization of the original SQUID coupler circuit of Fig.~\ref{fig:SQUIDCoupler} and the gradiometric SQUID coupler is given by $\Phi_{e,1} = \Phi_{e,i}$ and $\Phi_{e,2} = (\Delta\Phi_{e,o} -\Phi_{e,i})/2$, with the original operating condition $\Phi_{e,1} + 2\Phi_{e,2}=0$ being equivalent to $\Delta\Phi_{e,o} = 0$. Sensitivities of the qubit eigenfrequencies to external fluxes at the operating condition in the gradiometric SQUID coupler are given by:

{\small
\begin{align}
\frac{\partial\omega_{|\tilde{ij}\rangle}}{\partial\varphi_{e,i}}|_{\Delta\Phi_{e,o}=0} = \frac{1}{2}\Sigma E_{J,C}\sin{\left(\frac{\varphi_{e,1}}{2}\right)}\langle\cos{\left(\hat{\varphi}_2 -\hat{\varphi}_1\right)}\rangle, \label{eq:extfluxderivatives_inner} \\ 
\frac{\partial\omega_{|\tilde{ij}\rangle}}{\partial\varphi_{e,o}}|_{\Delta\Phi_{e,o}=0}=  \frac{1}{2}\Delta E_{J,C}\sin{\left(\frac{\varphi_{e,1}}{2}\right)}\langle\cos{\left(\hat{\varphi}_2 -\hat{\varphi}_1\right)}\rangle, \label{eq:extfluxderivatives_outer} \\ 
\frac{\partial\omega_{|\tilde{ij}\rangle}}{\partial\varphi_{e,o}'}|_{\Delta\Phi_{e,o}=0}= -\frac{1}{2}\Delta E_{J,C}\sin{\left(\frac{\varphi_{e,1}}{2}\right)}\langle\cos{\left(\hat{\varphi}_2 -\hat{\varphi}_1\right)}\rangle,
    \label{eq:extfluxderivatives_outerprime}
\end{align}
}

\noindent where we have used the approximation $\langle\sin{(\varphi_{2}-\varphi_1)}\rangle \approx 0$. As can be seen from eqs.~(\ref{eq:extfluxderivatives_inner}-\ref{eq:extfluxderivatives_outerprime}), the sensitivity to inner SQUID flux noise scales with $\Sigma E_{J,C}$, while the sensitivity to flux noise in the outer SQUID loops scales with asymmetry in the coupler junctions, $\Delta E_{J,C}$. 

We begin with an assessment of the impact of noise in the outer SQUID loops. As the loop area and perimeter of these outer loops can be relatively large they can host a significant amount of flux noise~\cite{braumuller2020squidgeometryperimeter}. In Fig.~\ref{fig:sensitivity}b we provide an estimate of the echo dephasing times in the qubit states $\ket{\tilde{10}}$ and $\ket{\tilde{01}}$ in the presence of flux noise as a function of $\Delta E_{J,C}$. In this analysis we consider the coupler to be biased at the operational point, ($\Phi_{e,1} = \Phi_\text{off}$), and we assume an external flux noise model consisting of $1/f$ noise with amplitude $A_{\Phi_{e,i}} = 10^{-6}\Phi_0$ for the smaller inner SQUID loop and $A_{\Phi_{e,o}} = A_{\Phi_{e,o}'}= 5\times10^{-6}\Phi_0$ for the larger outer SQUID loops, consistent with typically measured flux noise in superconducting circuits~\cite{braumuller2020squidgeometryperimeter} and for a circuit layout similar to that proposed below in Sec.~\ref{Section:Tiling}. The echo dephasing time is calculated as per Ref.~\cite{bylander2011noise}:
{\small
\begin{align}
    \frac{1}{T_{\phi, \text{echo}}^{1/f, \ket{ij}}} = \sqrt{\ln{2}}\sqrt{A_{\Phi_{e,i}}^2\left|\frac{\partial\omega_{\ket{\tilde{ij}}}}{\partial\Phi_{e,i}}\right|^2 + A_{\Phi_{e,o}}^2\left|\frac{\partial\omega_{\ket{\tilde{ij}}}}{\partial\Phi_{e,o}} - \frac{\partial\omega_{\ket{\tilde{ij}}}}{\partial\Phi_{e,o}'}\right|^2}, \label{eq:echodephasingtime}
\end{align}
}

\noindent where we take the inner and outer SQUID flux noises to be uncorrelated, but assume a worst-case correlation $\delta\Phi_{e,o}' = -\delta\Phi_{e,o}$ between the noise in the outer SQUID fluxes. This analysis shows that even for junction asymmetries of $\Delta E_{J,C}/2\pi = 160$ MHz (20\%) we find coherence times $T_{\phi, \text{echo}} > 160$~$\mu$s, indicating that for realistic noise and experimentally achievable junction fabrication and circuit layouts we don't anticipate substantial degradation in qubit coherence. 

Having examined the leading contribution to qubit dephasing associated with junction asymmetry, we now set $\Delta E_{J,C} = 0$ and discuss the contribution to dephasing arising from $\Sigma E_{J, C}$. In this case, since the cross-Kerr interaction rate and flux sensitivity originate from the same diagonal term proportional to $\Sigma E_{J,C}$, and $\Phi_\text{off}$ remains nearly unaffected, a clear trade-off exists between flux-noise-induced dephasing and gate speed. We analyze this trade-off by comparing the echo dephasing times and the minimum achievable gate time, $\min{T_{G}} = \pi/|\zeta_\text{on}|$.

We use the method presented in Ref.~\cite{rosenfeld2024fluxonium} to simulate the impact of flux noise on the CZ gate error. As the maximum ZZ interaction occurs at the sweet spot, $\Phi_{e,i} = \Delta\Phi_{e,o} = 0$, we expect the entangling phase of the CZ gate to be robust against flux noise. Figure~\ref{fig:sensitivity}d shows the estimated coherent error of the 22 ns-long CZ gate as a function of static offsets $\delta\Phi_{e,i}$ and $\delta(\Delta\Phi_{e,o})$ added to the flux waveform. In this analysis we use local $\hat{Z}$ phases found from the zero offsets to simulate drifts in local phases due to flux noise. The green and gray dashed curves calibrate the zero offsets to root-mean-square (RMS) flux deviations in a 1-hour drift period for $1/f$ flux noise of amplitudes of $A_{\Phi_{e,i}} = A_{\Phi_{e,o}} = 5\times10^{-6}\Phi_0$ and an anti-correlation between the two outer SQUID flux noises. From this analysis we find that the CZ gate maintains a small coherent error below the $10^{-6}$ level over a 1-hour drift period. 

\section{Unconventional crosstalk}
\label{Section:Crosstalk}

\begin{figure}[tbp]
\centering
\includegraphics[width = \columnwidth]{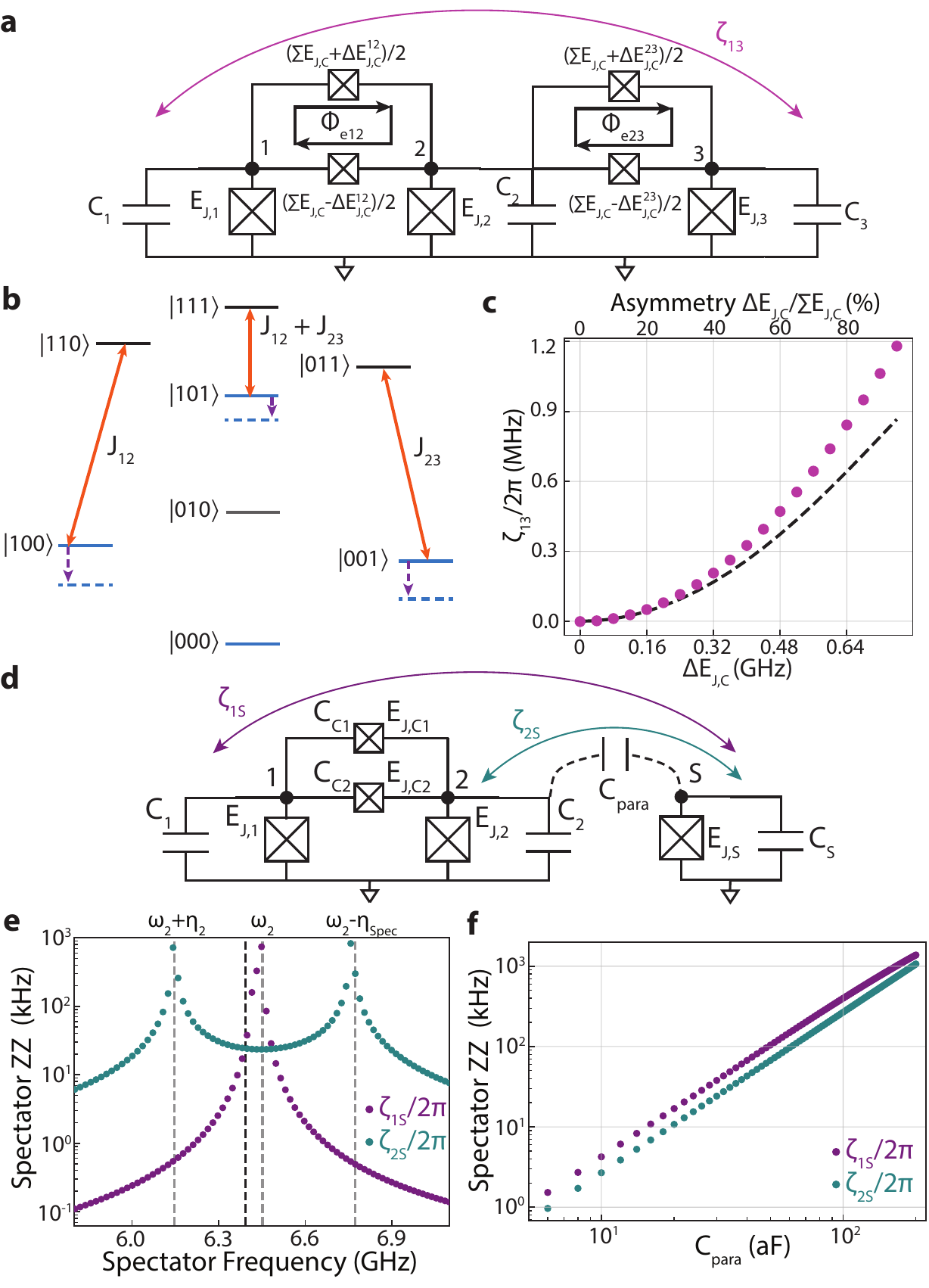}
\caption{\textbf{Unconventional crosstalk mediated by SQUID couplers.} \textbf{a}, Circuit schematic of a chain of three transmons with nearest-neighbor connection via SQUID couplers. Coupler capacitances are omitted for brevity. \textbf{b}, Energy level diagram of \textbf{a} with transition matrix elements (solid orange lines) due to longitudinal interaction provided in eq.~(\ref{eq:Hlongitudinal}). \textbf{c}, ZZ interaction rate $\zeta_{13}$ calculated from circuit quantization numerically as a function of $\Delta E_{J,C}$ (purple circles). Black dashed line indicates $\zeta_{13}$ obtained from perturbation theory. \textbf{d}, Circuit schematic of two transmons coupled via a SQUID coupler and a spectator transmon coupled to transmon 2 over a parasitic capacitance $C_\text{para}$. \textbf{e}, Spectator ZZ interaction rates $\zeta_{1S}$ and $\zeta_{2S}$ as functions of spectator 0-1 transition frequency $\omega_S$, with $C_\text{para} = 30 \ \text{aF}$. \textbf{f}, $\zeta_{1S}$ and $\zeta_{2S}$ as functions of $C_\text{para}$, estimated at $\omega_S/2\pi = \omega_{2}/2\pi - 60\ \text{MHz}$ (black dashed line in \textbf{e}). 
}
\label{Fig:Crosstalk}
\end{figure}

In an array of coupled qubits, we find that junction asymmetries in the SQUID coupler can induce crosstalk between next-nearest-neighbor pairs of qubits, as illustrated in Fig.~\ref{Fig:Crosstalk}a. This crosstalk results from odd-parity interactions, given by $\hat{H}_{\text{asym}-\text{asym}}$:

\begin{align}
    &\hat{H}_{\text{asym}-\text{asym}} = \Delta E_{J,C}^{12}\sin{\left(\frac{\varphi_{e12}}{2}\right)} \sin{(\hat{\varphi}_2-\hat{\varphi}_1)} \notag\\& \ \ \ \ \ \ \ + \Delta E_{J,C}^{23}\sin{\left(\frac{\varphi_{e23}}{2}\right)} \sin{(\hat{\varphi}_3-\hat{\varphi}_2)},
    \label{eq:Hasymasym}
\end{align}

\noindent where $\Delta E_{J,C}^{12}$ and $\Delta E_{J,C}^{23}$ are differential Josephson energies due to junction asymmetry within the SQUID couplers which directly connect the nearest-neighbor pair of qubits and the next-nearest-neighbor pair of qubits, respectively, and $\varphi_{e12}$ and $\varphi_{e23}$ are are the corresponding external fluxes threading the SQUID couplers. 

Expansion of eq.~(\ref{eq:Hasymasym}) to the third order reveals longitudinal interaction:

\begin{align}
    &\hat{H}_{\text{longitudinal}} \approx \left(J_{12} \hat{a}_{1}^\dagger\hat{a}_1 + J_{23} \hat{a}_{3}^\dagger\hat{a}_3\right)(\hat{a}_2 + \hat{a}_2^\dagger),
    \label{eq:Hlongitudinal}
\end{align}

\noindent where $J_{12} \approx \Delta E_{J,C}^{12'} (\varphi_{1}^{zpf})^2\varphi_{2}^{zpf}\sin{(\varphi_{e12}/2)}$ and $J_{23} \approx -\Delta E_{J,C}^{23'} (\varphi_{3}^{zpf})^2\varphi_{2}^{zpf}\sin{(\varphi_{e23}/2)}$. This is equivalent to a $Z_1/Z_3$-parity-dependent driving of transmon 2, which for detuned qubits realizes an effective ZZ interaction between transmon 1 and 3, similar to the gate scheme described in Ref.~\cite{gorshkov2025cavitymediatedcrosscrossresonancegate}. The relevant transitions are illustrated in Fig.~\ref{Fig:Crosstalk}b. The ZZ interaction rate $\zeta_{13}$ obtained from second-order perturbation theory is given by:

\begin{align}
    \zeta_{13} &\approx \frac{(J_{12}+J_{23})^2}{-\omega_2} - \frac{J_{12}^2}{-\omega_2} - \frac{J_{23}^2}{-\omega_2} = \frac{-2J_{12}J_{23}}{\omega_2}.
    \label{eq:ZZ13}
\end{align}

\noindent Using circuit quantization, $\zeta_{13}$ is calculated numerically as a function of $\Delta E_{J,C}^{12}=\Delta E_{J,C}^{23}=\Delta E_{J,C}$, and plotted in Fig.~\ref{Fig:Crosstalk}c. At each $\Delta E_{J,C}$, the junction capacitances are updated to maintain plasma frequencies and $\zeta_{13}$ is estimated under the idle conditions $\zeta_{12} = \zeta_{23}=0$. The perturbative prediction of eq.~(\ref{eq:ZZ13}) is in good agreement with the numerical circuit quantization result, indicating that most of $\zeta_{13}$ originates from the longitudinal interaction. Note that for $\Sigma E_{J,C}/2\pi = 0.8$ GHz, a junction asymmetry less than 20\% is sufficient to suppress $\zeta_{13}/2\pi$ below 60 kHz. 

Due to the weak dependence of the cross-Kerr interaction mediated by junction couplers, one can not rely on qubit frequency dispersion to suppress unwanted interactions within an array of coupled qubits. As such, crosstalk in the context of interactions with spectator qubits is a potential limitation of any SQUID-coupler-based architecture~\cite{cai2021crspectator, krinner2020bmarkingspectator, Jurcevic2022spectatordephasing}. As illustrated in the circuit shown in Fig.~\ref{Fig:Crosstalk}d, SQUID couplers can propagate cross-Kerr interaction from a qubit that is not directly coupled to a spectator qubit. In this circuit, transmon 1 and 2 are directly connected via a SQUID coupler, and a spectator transmon of frequency $\omega_\text{S}$ (subscripted ``S") couples to transmon 2 with a rate $g_\text{para}$ through a parasitic capacitance $C_\text{para}$. This coupling hybridizes transmon 2 and the spectator qubit: $\hat{a}_2 \xrightarrow{} \sqrt{1 - |\gamma|^2}\hat{a}_2 + \gamma\hat{a}_S$, where $\gamma \approx g_\text{para}/(\omega_2 - \omega_\text{S})$. Following eq.~(\ref{eq:Hint_flux_removed}), this parasitic hybridization is translated into ZZ interaction $\zeta_{1S}$ between transmon 1 and the spectator as follows: 
{\small
\begin{equation}
    \zeta_{1S} \approx -\Sigma E_{J,C}'\cos{\left(\frac{\varphi_{e,1}}{2}\right)}(\varphi_1^{zpf})^2 (\varphi_2^{zpf})^2 |\gamma|^2 \approx\zeta^{(1)}|\gamma|^2.
\end{equation}
}

\noindent This ``indirect" ZZ crosstalk is maximized when the two transmons 1 and 2 exhibit maximal ZZ interaction, i.e., during CZ gate operations. As shown in Fig.~\ref{Fig:Crosstalk}e, we calculate $\zeta_{1S}$ as a function of $\omega_S$ at $\varphi_{e,1} = 0$, for parameters listed in Table. \ref{Table:circuitparams}, $C_S = 69.2$ fF, and $C_\text{para} = 30$ aF. $\omega_S$ is varied by changing $E_{J,S}$. We also provide a comparison with conventional ``direct" ZZ crosstalk $\zeta_{2S}$. The indirect ZZ crosstalk is maximized near the resonance condition $\omega_S \approx \omega_2$ and can even be stronger than $\zeta_{2S}$, as illustrated in Fig.~\ref{Fig:Crosstalk}e and f. This adds emphasis to avoiding resonant frequency collision $\omega_S \approx \omega_2$ when allocating qubit frequencies \cite{brink2018freqcollision, morvan2022freqalloc}, and motivates the suppression of stray capacitances to levels at or below tens of aF (see Fig.~\ref{Fig:Crosstalk}f). 

\section{Fully miniaturized mergemon architecture}
\label{Section:Tiling}

\begin{figure}[tbp]
\centering
\includegraphics[width = \columnwidth]{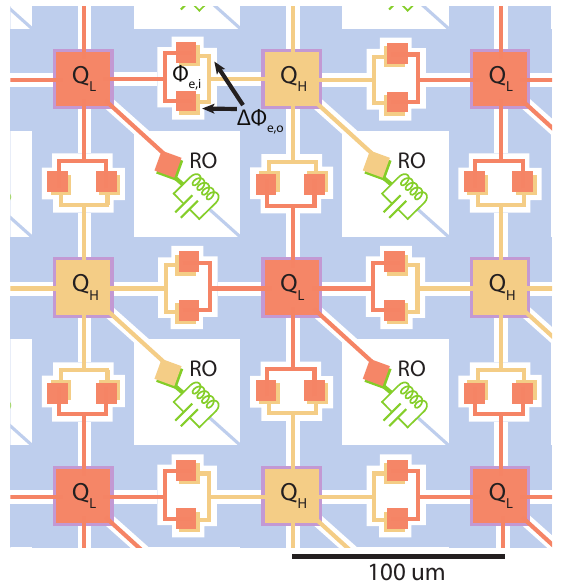}
\caption{\textbf{Mergemons on a square lattice for a miniaturized quantum processor.} Orange (yellow) shape corresponds to high voltage side metalization for low frequency (high frequency) qubits, where an overlap with low voltage side metalization (purple) forms a mergemon. High voltage side metalization of each mergemon extends to their nearest-neighbors, forming SQUID couplers. Low voltage side metalization is connected galvanically to ground metalization (blue), forming the outer SQUID loops. Each mergemon is connected galvanically to its lumped-element readout resonator (green) via a Josephson junction for junction readout. For convenience in illustration, elements including Josephson junctions are intentionally displayed bigger than their realistic relative length scales. Scale bar represents the anticipated lattice constant provided in the main text.
}
\label{fig:Tiling}
\end{figure}

In this section we present a qubit tiling strategy towards a fully miniaturized superconducting quantum processor based on mergemon qubits coupled together with SQUID couplers. A schematic illustration of the proposed circuit is shown in Fig.~\ref{fig:Tiling}. In this architecture all qubits and coupling elements are implemented using Josephson junctions, and free from bulky capacitor elements, greatly increasing the possible density of qubits within the array. It is envisioned that the same oxide barrier thickness for SQUID couplers and high-frequency mergemons can be employed, maintaining high CZ gate performance and reducing the complexity of the circuit fabrication (see Appendix \ref{App:circuitparameters}). Note that the same tiling strategy can also be used for transmon qubits with dedicated shunt capacitors, simply by enlarging the extended leads to serve as capacitor pads. 

For the circuit layout proposed in Fig.~\ref{fig:Tiling} there is a common ground (blue) connected to all of the low-voltage-side bottom junction electrodes (purple) of each mergemon qubit. The high-voltage-side top junction electrodes (orange and yellow) of the mergemon qubits have four lead extensions that connect each mergemon qubit to each of its four nearest-neighbors via a SQUID coupler. The low-voltage-side junction electrodes of the mergemon qubits, being galvanically connected to the common ground layer, give rise to the outer SQUID loops discussed in previous sections. The two junction electrode layers of each SQUID coupler separate the mergemon qubits in the square lattice into alternating ``high" (H) and ``low" (L) groupings, where the mergemon qubits labelled $Q_{L}$ (colored orange) have their top junction electrodes coupled to the upper metalization layer of each junction-based coupler, and the mergemon qubits labelled $Q_H$ (colored yellow) have their top junction electrodes connected to the lower junction electrode of the couplers. 

This separation of metallization layers avoids hop-over wiring elements and limits the extraneous superconducting loops in the lattice of qubits. The regular square lattice layout also gives rise to a gradiometric design that makes the circuit insensitive to homogeneous magnetic flux~\cite{Schwarz2013gradiometricfluxqubit,braumuller2016concentric,hyyppa2022unimon, bright2025gradiometric}. For maximal control, two flux delivery lines per SQUID coupler are required (control lines not shown). An alternative is to use a flux delivery line that satisfies the operating condition $\Delta\Phi_{e,o} = 0$.  

In this architecture, qubit frequencies can be dispersed over a relatively large range thanks to the strong cross-Kerr coupling provided by the SQUID coupler even under large qubit detunings. For instance, data qubits can be assigned to a low frequency band around $4.5$ GHz and auxiliary qubits can be assigned to a high frequency band around $6.3$ GHz. This mitigates frequency collision between nearest-neighbor pairs, enables the use of fixed-frequency qubits, and maximizes adiabaticity in CZ gates. To avoid accidental collision with strongly-coupled TLS defects, flux-tunable transmons may also be used \cite{zhao2020mergemon, mamin2021mergemondesign, daum2025investigationparasitictwolevelsystems}.

As illustrated in Fig.~\ref{fig:Tiling}, the top junction electrode of each mergemon qubit is galvanically connected to a lumped-element read-out resonator via a Josephson junction for qubit state read-out. This junction-based readout technique not only eliminates coupling capacitors for readout but can also simplify the read-out circuit by obviating the need for an additional Purcell filter~\cite{chapple2025balanced, wang2025longitudinal}. Moreover, the SQUIDs introduced by junction readout can provide tunability in nonlinear interactions and effective linear coupling, as demonstrated in Ref.~\cite{wang2025longitudinal}. We anticipate that this tunability can be used for unconditional qubit reset or leakage reduction by activating hybridization to readout resonators \cite{wang2025longitudinal,reed2010fast, magnard2018fast, mcewen2021removing, zhou2021rapid, marques2023allmicrowaveleakage, lacroix2023fast, kim2025unconditional}. 

The use of lumped-element readout resonators with compact footprints~\cite{zotova2023compactppc,bell2012superinductor, long2019fluxonium, pechenezhskiy2020blochmonium, junger2025superinductor} further enables the miniaturized tiling of the qubit lattice. In addition, by using a flip-chip architecture~\cite{conner2021flipchip, Kosen2022buildingblocks}, charge drive lines, flux delivery lines, and readout feedlines can be placed on an opposing ``wiring" chip without consuming space on the qubit chip. Combining these features together, we estimate that a miniaturized mergemon processor with a qubit lattice constant of $100$ um is achievable, enabling the packing of up to $10^6$ qubits in an area of $10$ cm $\times$ $10$ cm that could be fabricated on a single 6-inch wafer~\cite{gidney2025factor2048bitrsa}. 

Looking forward, there are myriad other fabrication and layout considerations that one can consider in such a mergemon-qubit architecture. One particularly appealing opportunity afforded by an all-junction architecture is the utilization of acoustic bandgap structures to engineer the phonon bath environment of the circuit elements. As proposed~\cite{behunin2016dimensional, gregory2020ultralongphonon} and recently demonstrated~\cite{chen2024phonon, odeh2025nonmarkovian}, by shutting off the primary phonon bath decoherence channel of two-level-system (TLS) defects one can dramatically improve the coherence properties of TLS, and correspondingly the coherence properties of electrical or acoustic elements that couple to the TLS. This effect is particularly dramatic for elements that have a small surface area such that only a discrete set of two-level-system (TLS) are spectrally nearby, such as in a Josephson junction. Other considerations, such as minimizing the energy participation in the wiring leads of the tiled mergemon qubits and reducing parasitic capacitance to spectator qubits, are described in Appendices \ref{App:interfacialloss} and \ref{App:parasiticcaps}.

\section{Other uses of the SQUID coupler}
\label{Section:OtherUtility}

In this section we consider a few alternative uses of a SQUID coupler with similar hierarchy of coupler junction energy to operation frequency as proposed for transmon or mergemon qubit architectures. One such interesting application is to the engineering of two-photon exchange interactions used to dissipatively stabilize cat qubits \cite{Mirrahimi2014firstcatqubit, leghtas2015confining, lescanne2020exponential, putterman2025hardware}. Under the SQUID coupler operating condition $\varphi_{e,1} + 2\varphi_{e,2} =\pi$, the odd-parity interaction $\hat{H}_\text{sine}$ is obtained:

\begin{align}
    \hat{H}_\text{sine} &=\Sigma E_{J,C}\cos{\left(\frac{\varphi_{e,1}}{2}\right)}\sin{(\hat{\varphi}_2-\hat{\varphi}_1)}. 
    \label{eq:Hint_sine}
\end{align}

\noindent By introducing a time-dependent drive to the SQUID loop flux, $\varphi_{e,1}(t) = \pi + \epsilon_d\cos{(\omega_d t)}$, with $\omega_d \approx 2\omega_{1} - \omega_2$ on $\varphi_{e,1}$, this yields under rotating-wave approximation a two-photon exchange interaction $\hat{H}_\text{two-photon}$ analogous to the result of Ref.~\cite{stolyarov2025twophoton}:

\begin{align}
    \hat{H}_\text{two-photon} \approx g_2\hat{a}_1^2\hat{a}_2^\dagger     + h.c., \label{eq:H2}
\end{align}

\noindent where $g_2 \equiv \frac{1}{8}\epsilon_d\Sigma E_{J,C}'(\varphi_{1}^{zpf})^2 (\varphi_{2}^{zpf})$. In the dissipatively-stabilized cat qubit system, mode 1 would be a harmonic oscillator storage mode and mode 2 could be a transmon qubit, for instance, used for ``buffering" the storage mode. Remarkably, the $g_2$ coupling rate is directly proportional to the bare zero-point fluctuations of the two modes rather than their hybridized values. This separates the strength of the two-photon dissipation from that of the self- and/or cross-Kerr induced on the storage mode by the nonlinear buffer circuit due to mode hybridization~\cite{marquet2024autoparametric, putterman2025preserving}. It also allows for more flexibility in allocating storage and buffer mode frequencies than in the conventional capacitively-coupled situation, where much larger detunings between elements may be employed. 


Another potential use of the proposed SQUID coupler lies in engineering next-nearest-neighbor qubit interactions. For the three-transmon circuit illustrated in Fig.~\ref{Fig:Crosstalk}c, we obtain odd-parity interactions for the operating condition $\varphi_{e,1} + 2\varphi_{e,2} =\pi$:

\begin{align}
    &\hat{H}_{\text{sine}-\text{sine}} = \Sigma E_{J,C}^{12}\cos{\left(\frac{\varphi_{e12}}{2}\right)} \sin{(\hat{\varphi}_2-\hat{\varphi}_1)} \notag\\& \ \ \ \ \ \ \ + \Sigma E_{J,C}^{23}\cos{\left(\frac{\varphi_{e23}}{2}\right)} \sin{(\hat{\varphi}_3-\hat{\varphi}_2)}.
    \label{eq:Hlongrange}
\end{align}

\noindent This interaction renders the same longitudinal interaction and ZZ interaction rate $\zeta_{13}$ between the next-nearest-neighbor pair provided in eq.~(\ref{eq:Hlongitudinal}-\ref{eq:ZZ13}). Notably, $\zeta_{13}/2\pi$ can be larger than 1 MHz for $\Sigma E_{J,C}/2\pi = 0.8$ GHz as shown in Fig.~\ref{Fig:Crosstalk}f, and can be strengthened quadratically by increasing $\Sigma E_{J,C}$s. 

Finally, the SQUID coupler can provide tunable XX interactions that can be used for the implementation of an adiabatic $\sqrt{\text{iSWAP}}$ gate between dual-rail transmons \cite{kubica2023erasure, levine2024dualrail}. This is derived from the fact that a single transmon Pauli-Z ($\hat{Z}$) operator is equivalent to a Pauli-X ($\hat{X}_{DR}$) operator of a dual-rail transmon that contains the transmon up to a global phase, as follows:
\begin{align}
    (\hat{I}_{1}\hat{Z}_{2})(\hat{Z}_{3}\hat{I}_{4}) = -\hat{X}_{DR,a}\hat{X}_{DR,b},
    \label{eq:ZZisXXinDualrail}
\end{align}
\noindent where subscript $i \in \{1,2,3,4\}$ denotes transmon $i$, and dual-rail transmon $a$ ($b$) consists of transmons 1, 2 (3, 4). Thus, if the SQUID coupler provides a tunable ZZ interaction in the form of $(\zeta/4) \hat{Z}_2\hat{Z}_3$ between transmons 2 and 3, this is equivalent to a tunable XX interaction $(-\zeta/4) \hat{X}_{DR,a}\hat{X}_{DR,b}$ for the dual-rail transmons $a$ and $b$. This allows for the implementation of a $\sqrt{\text{iSWAP}}$ gate with duration $T_G^\text{min} = \pi/|\zeta|$ plus an adiabaticity overhead \cite{kubica2023erasure}, which is as short as a CZ gate between transmons. In the context of dual-rail qubits and erasure detection, the SQUID coupler preserves the error hierarchy by not introducing additional channels for leakage errors \cite{kubica2023erasure,  levine2024dualrail, chou2024dualrail}. Furthermore, the impact of frequency noise is suppressed by operation at the artificial sweet-spot \cite{kubica2023erasure, levine2024dualrail}. See Appendix \ref{App:tiling} for the tiling of dual-rail transmons.

\section{Conclusion}
\label{Section:Conclusion}

The SQUID coupler provides modeless and tunable cross-Kerr coupling, which realizes a fast and high-fidelity CZ gate between transmons with minimal adiabaticity overhead. Sensitivities to junction asymmetry and flux noise are suppressed by choosing high qubit frequencies and relatively small junction energies. Unconventional crosstalk due to parasitic hybridization to spectator qubits and junction asymmetry are also shown to be sufficiently small for realistic circuit parameters. SQUID couplers are particularly interesting in the context of a mergemon qubit architecture, where junctions with small Josephson energies and reduced SQUID loop sizes due to the absence of bulky shunt capacitors are anticipated. Using the junction-based coupling schemes, we propose a scalable tiling strategy towards a fully miniaturized superconducting quantum processor. 

\begin{acknowledgments}
We thank Aashish A. Clerk for thoughtful review and constructive comments on the manuscript. We thank John Parker, Benjamin Boone, Lucia De Rose, Piero Chiappina, David P. Lake, Utku Hatipoglu, Sameer Sonar, Matt Davidson, Olivia Pitcl, Eunjong Kim, Arbel Haim, Ziwen Huang, Kyungjoo Noh, Yufeng Ye, Yoni Schattner, and Connor T. Hann for helpful discussions. This work was supported through a sponsored research agreement with Amazon Web Services. 
\end{acknowledgments}


\appendix

\renewcommand{\thefigure}{A\arabic{figure}}
\setcounter{figure}{0}

\section{Circuit Quantization}
\label{App:circuitquantization}

The SQUID coupler circuit is analyzed following the standard circuit quantization procedure \cite{vool2017circuitqedintro, you2019circuitquantization, riwar2022circuitquantization}. We define external flux drops $\Phi_{e, \text{top}}$ ($\Phi_{e, \text{bot}}$) as the external flux drop across the top (bottom) Josephson junction of the SQUID coupler, indicated by the blue (red) arrow, and $\Phi_{e,Ji}$s as the external flux drop across the transmon junctions. The branch fluxes $\Phi_{J1}$, $\Phi_{J2}$ across Josephson junctions of transmon 1 and 2, and $\Phi_\text{top}$ and $\Phi_\text{bot}$ across the top and bottom Josephson junctions of the SQUID coupler are represented as follows:


\begin{align}
    \Phi_{J1} &=\phi_1 + \Phi_{e, J1}, \ \Phi_{J2} = -\phi_2 + \Phi_{e,J2},\notag \\\Phi_\text{bot} &=  \phi_2 - \phi_1 + \Phi_{e, \text{bot}}, \ \Phi_\text{top} = \phi_2 - \phi_1 + \Phi_{e, \text{top}}, \notag \\
    \Phi_{e,2} &= \Phi_{e, J1} + \Phi_{e,\text{bot}} + \Phi_{e, J2}, \notag\\
    \Phi_{e,1} &= \Phi_{e, \text{top}} - \Phi_{e, \text{bot}}, 
\end{align}

\noindent where $\phi_1$, $\phi_2$ are node variables of the two transmons.

\begin{figure}[tbp]
\centering
\includegraphics[width = \columnwidth]{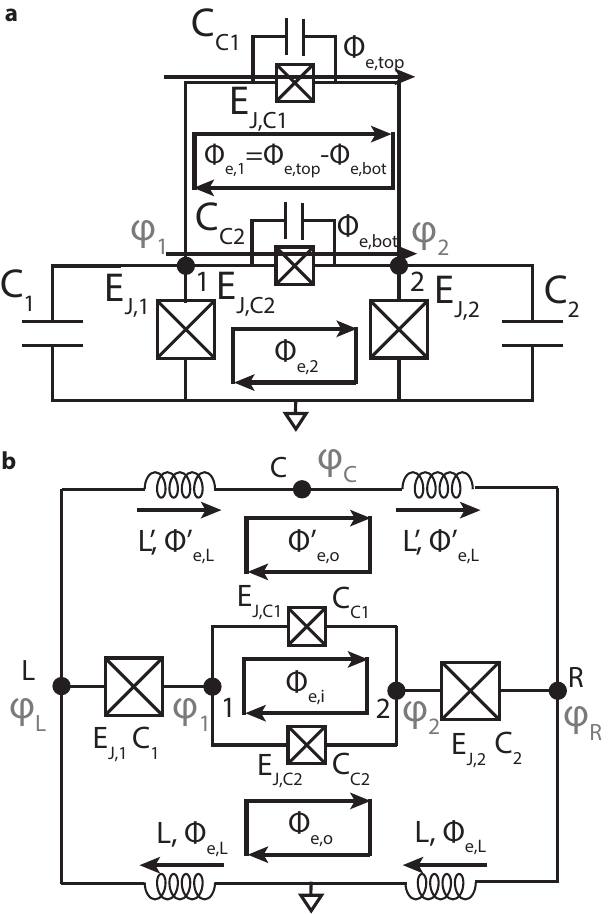}
\caption{\textbf{SQUID coupler with two transmons.} \textbf{a}, Single-sided SQUID coupler with two transmons. \textbf{b}, SQUID coupler with small inductances around ground connection, defining physical SQUID loops.}
\label{Fig:SQUIDCouplerCircuit}
\end{figure}

\noindent The Lagrangian of the circuit is given by:
{
\small
\begin{align}
    \mathcal{L}&=\frac{1}{2}C_1 (\dot{\phi}_{1}+\dot{\Phi}_{e, J1})^2 + \frac{1}{2}C_2 (\dot{\phi}_{2}-\dot{\Phi}_{e, J2} )^2 \notag\\ &+ \frac{1}{2}C_{C1} (\dot{\phi}_{2}-\dot{\phi}_{1} + \dot{\Phi}_{e, \text{top}})^2 + \frac{1}{2}C_{C2} (\dot{\phi}_{2}-\dot{\phi}_{1} + \dot{\Phi}_{e,\text{bot}})^2  \notag\\+&E_{J,1}\cos{\left(2\pi\frac{(\phi_{1} + \Phi_{e,J1})} {\Phi_0}\right)}+ E_{J,2}\cos{\left(2\pi\frac{(\phi_{2} -\Phi_{e,J2})} {\Phi_0}\right)}\notag\\ &+E_{J,C1}\cos{\left(2\pi\frac{(\phi_{2} - \phi_{1}  + \Phi_{e, \text{top}})}{\Phi_0}\right)} \notag \\&+ E_{J,C2}\cos{\left(2\pi\frac{(\phi_{2} - \phi_{1} + \Phi_{e, \text{bot}})}{\Phi_0}\right)},
    \label{eq:lagrangian}
\end{align}
}

\noindent and the Legendre transformation:
{
\small
\begin{align}
Q_1 &= \frac{\partial\mathcal{L}}{\partial\dot{\phi}_1}  = C_1 (\dot{\phi}_1 + \dot{\Phi}_{e, J1}) + C_{C} (\dot{\phi}_1 - \dot{\phi}_2 - \dot{\Phi}_\text{e,12}), \notag\\ 
Q_2 &= \frac{\partial\mathcal{L}}{\partial\dot{\phi}_2}  = C_2 (\dot{\phi}_2 - \dot{\Phi}_{e, J2}) + C_{C} (\dot{\phi}_2 - \dot{\phi}_1 + \dot{\Phi}_\text{e,12}),
\label{eq:legendretransform}
\end{align}
}

\noindent provides the Hamiltonian:
{
\small
\begin{align}
    &H = Q_1\dot{\phi}_{1} + Q_2\dot{\phi}_{2} - \mathcal{L} \notag\\ &= \frac{C_2 + C_C}{2C^2}Q_1^2  + \frac{C_1 + C_C}{2C^2}Q_2^2 + \frac{C_C}{C^2}Q_1Q_2 \notag\\  & + Q_1 \left(\frac{C_2C_C\dot{\Phi}_\text{e,12} - C_1(C_2+C_C)\dot{\Phi}_{e, J1} + C_2C_C\dot{\Phi}_{e, J2}}{C^2} \right)\notag\\ &+  Q_2  \left(\frac{-C_1C_C\dot{\Phi}_\text{e,12} - C_1C_C\dot{\Phi}_{e, J1} + C_2(C_1+C_C)\dot{\Phi}_{e, J2}}{C^2} \right)\notag  \\ &-E_{J,1}\cos{\left(2\pi\frac{(\phi_{1} + \Phi_{e,J1})} {\Phi_0}\right)} -E_{J,2}\cos{\left(2\pi\frac{(\phi_{2} - \Phi_{e,J2})} {\Phi_0}\right)}\notag\\ &-E_{J,C1}\cos{\left(2\pi\frac{(\phi_{2} - \phi_{1}  + \Phi_{e,\text{top}})}{\Phi_0}\right)} \notag \\&- E_{J,C2}\cos{\left(2\pi\frac{(\phi_{2} - \phi_{1} + \Phi_{e,\text{bot}})}{\Phi_0}\right)},
    \label{eq:hamiltonian}
\end{align}
}

\noindent where we have introduced the following definitions for a succinct description,
\begin{align}
&C_C \equiv C_{C1} + C_{C2},\notag\\ 
\Phi_{e, 12} \equiv &\frac{C_{C1}}{C_C} \Phi_{e, \text{top}} + \frac{C_{C2} }{C_C}\Phi_{e, \text{bot}},\notag\\
C^2 \equiv &C_1C_2 + C_1C_C + C_2C_C.
\end{align}

The irrotational constraint can be found by eliminating the terms containing time-derivatives of the external fluxes:
{
\begin{align}
    C_2C_C\dot{\Phi}_\text{e,12} - C_1(C_2 + C_C)\dot{\Phi}_{e, J1} + C_2C_C\dot{\Phi}_{e, J2} = 0, \notag \\ 
    -C_1C_C\dot{\Phi}_\text{e,12} - C_1C_C\dot{\Phi}_{e, J1} + C_2(C_1 + C_C)\dot{\Phi}_{e, J2} = 0,
\end{align}
}

\noindent which is satisfied with the following voltage-divider parameterization:
\begin{align}
    C_1\dot{\Phi}_{e, J1} = C_2\dot{\Phi}_{e, J2} = C_C\dot{\Phi}_{e, 12}.
    \label{eq:voltagedivider}
\end{align}

This reveals the relationship between the branch external flux drops and external fluxes by solving the linear equations and integrating over time:
\begin{align}
    &\Phi_\text{e, top} = \Phi_{e,12} + \frac{1-d_C}{2}
    \Phi_{e,1},\notag \\ 
    &\Phi_\text{e, bot} = \Phi_{e,12} -\frac{d_C+1}{2}
    \Phi_{e,1},\notag \\ 
    &\Phi_{e,2} = \Phi_{e,J1} + \Phi_{e,J2} + \Phi_{e,\text{bot}}, \notag\\ 
    =  C_C&\left(\frac{1}{C_1} + \frac{1}{C_2} + \frac{1}{C_C}\right)\Phi_{e, 12} - \frac{d_C+1}{2}
    \Phi_{e,1},
\end{align}

\begin{align}
    &\Phi_{e,12} = \frac{C_1C_2}{2C^2}\left((d_C+1){\Phi}_{e,1} + 2{\Phi}_{e,2}\right),\notag\\
    &\Phi_{e, J1} = \frac{C_2C_C}{2C^2}\left((d_C+1){\Phi}_{e,1} + 2{\Phi}_{e,2}\right),\notag\\
    &\Phi_{e, J2} = \frac{C_1C_C}{2C^2}\left((d_C+1){\Phi}_{e,1} + 2{\Phi}_{e,2}\right),\notag \\ 
   &\Phi_{e, \text{top}} = \Phi_{e, \text{bot}} + \Phi_{e, 1}, \notag \\ 
   \Phi_{e, \text{bot}} = &\left(\frac{C_1C_2}{C^2} - 1\right)\frac{(d_C+1)}{2}\Phi_{e,1} + \frac{C_1C_2}{C^2}\Phi_{e,2}, 
\end{align}
\noindent where $d_C \equiv (C_{C1} - C_{C2})/C_C$ is the junction capacitance asymmetry in the SQUID coupler. Note that if both junction capacitances and the Josephson energies are mostly proportional to the junction area, it is reasonable to assume that the Josephson energy asymmetry and the capacitance asymmetry are the same.

We can examine the consistency of the solution by considering the extreme cases. At $C_{C2} \xrightarrow{} \infty$ (the bottom junction is replaced with a short), we find $\Phi_{e, \text{bot}} \xrightarrow{} 0$ and $\Phi_{e, \text{top}} \xrightarrow{} \Phi_{e, 1}$, consistent with a shorted single junction. At $C_1\xrightarrow{} 0$ (transmon 1's junction is replaced with a pure inductor or open), we find $\Phi_{e, \text{bot}} \xrightarrow{} -C_{C1}\Phi_{e, 1}/C_C$, $\Phi_{e, \text{top}} \xrightarrow{} C_{C2}\Phi_{e, 1}/{C_{C}}$, and $\Phi_{e, J1} \xrightarrow{} \Phi_{e,2}$, consistent with the DC SQUID and the fluxonium example provided in You et al.~\cite{you2019circuitquantization}. Finally, if we remove the top junction in the SQUID coupler, $C_{C1} \xrightarrow{} 0$, we recover the solution found in Appendix A of Campbell et al.~\cite{campbell2023modular}.

The complete expression for the Hamiltonian is as follows:
{
\small
\begin{align}
    &\hat{H}  = \notag\\ & \ \ \ \ 4E_{C,1}\hat{n}_1^2 -E_{J,1}\cos{\left(\hat{\varphi}_1 + \frac{C_2C_C}{2C^2}((d_C+1)\varphi_{e,1} +2 \varphi_{e,2})\right)}\notag\\
     & \ \ \ \ 4E_{C,2}\hat{n}_2^2 -E_{J,2}\cos{\left(\hat{\varphi}_2 - \frac{C_1C_C}{2C^2}((d_C+1)\varphi_{e,1} +2 \varphi_{e,2})\right)}\notag\\
     & \ \ \ \ -E_{J,C1}\cos{\left(\hat{\varphi}_2 - \hat{\varphi}_1 + \varphi_{e, \text{top}}\right)} \notag\\ & \ \ \ \ -E_{J,C2}\cos{\left(\hat{\varphi}_2 - \hat{\varphi}_1 + \varphi_{e, \text{bot}}\right)}
     +g\hat{n}_1\hat{n}_2,
    \label{eq:hamiltonian_full}
\end{align}
}

\noindent where node variables are replaced by operators (variables with hats). $\hat{\varphi}_i \equiv \frac{2\pi}{\Phi_0} \hat\phi_i$s are reduced flux operators, and $\hat{n}_i \equiv \frac{\hat{Q}_i}{2e}$ are reduced charge operators. $E_{C, 1}\equiv e^2 (C_2+C_C)/2C^2$, $E_{C, 2}\equiv e^2(C_1+C_C)/2C^2$ are the charging energy of the two modes,  and $g \equiv 4e^2C_C/C^2$ is the charge coupling rate. External fluxes are normalized to $\varphi_{e, k} \equiv 2\pi\Phi_{e,k}/\Phi_0 \ (k \in \{1,2\})$.

Under the symmetric SQUID approximation $d_C \approx 0$ and small junction approximation $C_C \ll C_1, C_2$ (note that the derivations up to eq.~(\ref{eq:hamiltonian_full}) are exact), $\varphi_\text{top}$ and $\varphi_\text{bot}$ approach $\varphi_{e,2}$ and $\varphi_{e,1} + \varphi_{e, 2}$, and we eliminate the external fluxes from the transmon junction potentials. As a result, we obtain the Hamiltonian provided in the main text:
\begin{align}
    \hat{H}  & \approx 4E_{C,1}\hat{n}_1^2 -E_{J,1}\cos{\left(\hat{\varphi}_1\right)}
     + 4E_{C,2}\hat{n}_2^2 -E_{J,2}\cos{\left(\hat{\varphi}_2\right)}\notag\\
     & -E_{J,C1}\cos{\left(\hat{\varphi}_2 - \hat{\varphi}_1 + \varphi_{e, 1}+\varphi_{e,2}\right)} \notag\\ &  -E_{J,C2}\cos{\left(\hat{\varphi}_2 - \hat{\varphi}_1 + \varphi_{e, 2}\right)}
     +g\hat{n}_1\hat{n}_2.
    \label{eq:hamiltonian_main_text}
\end{align}

\noindent Bringing back the asymmetry and the $C_C$ terms, the ideal external flux constraint $\varphi_{e,1} = -2\varphi_{e,2}$ produces the following full-circuit Hamiltonian without the approximations:
{
\small
\begin{align}
    &\hat{H}|_{\varphi_{e,2} = -\varphi_{e,1}/2}  = \notag\\ & \ \ \ \ 4E_{C,1}\hat{n}_1^2 -E_{J,1}\cos{\left(\hat{\varphi}_1 - \frac{C_2C_C}{2C^2}d_C\varphi_{e,1}\right)}\notag\\
     & \ \ \ \ 4E_{C,2}\hat{n}_2^2 -E_{J,2}\cos{\left(\hat{\varphi}_2 + \frac{C_1C_C}{2C^2}d_C\varphi_{e,1}\right)} +g\hat{n}_1\hat{n}_2\notag\\
     & \ \ \ \ -\Sigma E_{J,C}\cos{\left(\frac{\varphi_{e1}}{2}\right)}\cos{\left(\hat{\varphi}_2 - \hat{\varphi}_1 + \frac{(C_1+C_2)C_C}{2C^2}d_C\varphi_{e, 1}\right)} \notag\\ & \ \ \ \ +\Delta E_{J,C}\sin{\left(\frac{\varphi_{e1}}{2}\right)}\sin{\left(\hat{\varphi}_2 - \hat{\varphi}_1 + \frac{(C_1+C_2)C_C}{2C^2}d_C\varphi_{e, 1}\right)}.
    \label{eq:hamiltonian_ext_contraint}
\end{align}
}

\noindent where $\Sigma E_{J,C} \equiv E_{J,C1} + E_{J,C2}$, $\Delta E_{J,C} \equiv E_{J,C1} - E_{J,C2} \approx d_C\Sigma E_{J,C}$, and trigonometric relations are used to group cosine and sine terms. Within the cosinusoidal terms that contain the node variables, the external flux terms are suppressed to the scale of $d_CC_C/C_1$ and $d_CC_C/C_2 $, which are approximately $0.002$ assuming a modest asymmetry of $d_C \approx 0.2$ and $C_C$ used in this work. This validates the perturbative analysis without including these terms.

For the SQUID coupler shown in Fig.~\ref{Fig:SQUIDCouplerCircuit}b, which reflects the realistic circuit shown in the tiling schematic Fig.~\ref{fig:Tiling}, circuit quantization should be aided by properly considering the equilibrium node phases set by the small inductances around the low-impedance outer loops. Nodes $L$ (left), $R$ (right), and $C$ (center), and the corresponding node fluxes $\phi_L$, $\phi_R$, and $\phi_C$, are introduced by the inductor network consisting of 4 inductors with inductances $L, L'\ll L_J$, where $L_J$ is the inductance of the Josephson junctions. In this circuit, we introduce three external fluxes $\Phi_{e,i}, \Phi_{e,o}$, and $\Phi_{e,o}'$ threading the three distinct SQUID loops.

Defining external flux drops at the inductors as $\Phi_{e,L}$ and $\Phi_{e,L}'$, the following set of equations is obtained due to fluxoid quantization:

\begin{align}
    2\Phi_{e,L}' - (\Phi_{e,J1} + \Phi_{e,\text{top}}+\Phi_{e,J2}) = \Phi_{e,o}',\notag \\ 
    2\Phi_{e,L} + (\Phi_{e,J1} + \Phi_{e,\text{bot}}+\Phi_{e,J2}) = \Phi_{e,o},\notag \\ 
    \Phi_{e,\text{top}} - \Phi_{e,\text{bot}} = \Phi_{e,i}, &\notag \\ 
    \xrightarrow{} 2\Phi_{e,L} + 2\Phi_{e,L}' = \Phi_{e,i} + \Phi_{e,o} + \Phi_{e,o}' \equiv \Sigma \Phi_e.&
\end{align}

\noindent As it is assumed that $L$ and $L'$ are small, the node phases $\phi_L$, $\phi_R$, and $\phi_b$ are localized around the equilibrium points $\phi_{L0}$, $\phi_{R0}$, and $\phi_{C0}$. The inductive energy is given by:

\begin{align}
    E_{\text{ind}} &= \frac{1}{2}E_L(\phi_L + \Phi_{e,L})^2 + \frac{1}{2}E_L(-\phi_R + \Phi_{e,L})^2 \notag \\ &+ \frac{1}{2}E_L'(\phi_C - \phi_L + \Phi_{e,L}')^2 + \frac{1}{2}E_L'(\phi_R - \phi_C + \Phi_{e,L}')^2,\label{eq:potentialinductors}
\end{align}

\noindent where $E_L \equiv (\Phi_0/2\pi)^2 /L$ and $E_L \equiv (\Phi_0/2\pi)^2 /L'$ are inductive energies of the inductors. The equilibrium points are found by minimizing eq.~(\ref{eq:potentialinductors}) through solving the following set of equations: 


\begin{align}
    \frac{\partial E_\text{ind}}{\partial\phi_C} &= E_L'(\phi_C - \phi_L + \Phi_{e,L}') -  E_L'(\phi_R - \phi_C + \Phi_{e,L}'),\notag\\
    \frac{\partial E_\text{ind}}{\partial\phi_L} &= E_L(\phi_L + \Phi_{e,L}) -  E_L'(\phi_C - \phi_L + \Phi_{e,L}'),\notag\\
    \frac{\partial E_\text{ind}}{\partial\phi_R} &= -E_L(-\phi_R + \Phi_{e,L}) +  E_L'(\phi_R - \phi_C + \Phi_{e,L}'),\notag\\ 
  \xrightarrow{} \phi_{L0} &= -\phi_{R0} = \frac{E_L'\Phi_{e,L}' - E_L\Phi_{e,L}}{E_L + E_L'}, \  \phi_{C0} =0.
\end{align}

The role of $\phi_{L0}$ and $\phi_{R0}$, compared to the single-sided SQUID coupler, is to shift the external flux drops on qubit 1 and 2 by $\Phi_{e,J1} \xrightarrow{} \Phi_{e,J1} - \phi_{L0}$ and $\Phi_{e,J2} \xrightarrow{} \Phi_{e,J2} + \phi_{R0}$. Thus, the irrotational constraint and the subsequent algebra yields:

\begin{align}
    C_1\dot{\tilde\Phi}_{e, J1} = C_2&\dot{\tilde\Phi}_{e, J2} = C_C\dot{\Phi}_{e, 12},\notag \\
    \tilde\Phi_{e,J1} \equiv \Phi_{e,J1} - \phi_{L0}, &\ \tilde\Phi_{e,J2} \equiv \Phi_{e,J2} + \phi_{R0},
    \label{eq:voltagedividershifted}
\end{align}

\begin{align}
    \Phi_{e,o} &= (\Phi_{e,J1}-\phi_{L0}) + (\Phi_{e,J2}+\phi_{R0}) + \Phi_{e,\text{bot}} \notag\\&+ 2\Phi_{e,L} + \phi_{L0} - \phi_{R0} \notag\\ 
    &= \tilde\Phi_{e,J1} + \tilde\Phi_{e,J2} + \Phi_{e,\text{bot}} + \frac{E_L'(2\Phi_{e,L} + 2\Phi_{e,L}')}{E_L+E_L'},\notag \\ 
    \xrightarrow{}&\tilde\Phi_{e,J1} + \tilde\Phi_{e,J2} + \Phi_{e,\text{bot}} = \Phi_{e,o} -\frac{E_L'}{E_L + E_L'}\Sigma \Phi_e. \label{eq:externalfluxrelations}
\end{align}

\noindent The last line in eq.~(\ref{eq:externalfluxrelations}) indicates that the solutions found for the single-sided SQUID coupler can be used as follows:

\begin{align}
    \Phi_{e,1} \equiv &\Phi_{e,i}, \ \ \Phi_{e,2} \equiv \Phi_{e,o} - \frac{E_L'}{E_L + E_L'}\Sigma\Phi_e,\notag \\
    \Phi_{e,12} &= \frac{C_1C_2}{2C^2}\left((d_C+1){\Phi}_{e,i} + 2{\Phi}_{e,2}\right),\notag\\
    \tilde\Phi_{e, J1} &= \frac{C_2C_C}{2C^2}\left((d_C+1){\Phi}_{e,1} + 2{\Phi}_{e,2}\right),\notag\\
    \tilde\Phi_{e, J2} &= \frac{C_1C_C}{2C^2}\left((d_C+1){\Phi}_{e,1} + 2{\Phi}_{e,2}\right),\notag \\ 
   &\Phi_{e, \text{top}} = \Phi_{e, \text{bot}} + \Phi_{e, 1}, \notag \\ 
   \Phi_{e, \text{bot}} = &\left(\frac{C_1C_2}{C^2} - 1\right)\frac{(d_C+1)}{2}\Phi_{e,1} + \frac{C_1C_2}{C^2}\Phi_{e,2}. 
\end{align}

\noindent $\Phi_{e,2}$ converges to $\Phi_{e,o}$ as the ratio of the outer loop inductances approaches 0 $L/L'\xrightarrow{} 0$ ($E_L'/E_L \xrightarrow{} 0$), consistent with the single-sided SQUID coupler. A special case emerges at the symmetric inductances $L = L' \xrightarrow{}\Phi_{e,2} = (\Delta\Phi_{e,o} - \Phi_{e,i})/2$, at which the circuit is sensitive only to the differential of the outer external fluxes $\Delta \Phi_{e,o} \equiv \Phi_{e,o} - \Phi_{e.o}'$, thus called a ``gradiometric SQUID coupler", and the inner external flux. 

Finally, the operating condition $\Phi_{e,1} + 2\Phi_{e,2} = 0$ can be re-expressed as follows, for the general case and the gradiometric case:

\begin{align}
    (1-\frac{E_L'}{E_L})\Phi_{e,i} + 2(&\Phi_{e,o} -\frac{E_L'}{E_L}\Phi_{e,o}') = 0 \notag, \\ 
    \xrightarrow{L=L'} \Delta &\Phi_{e,o} = 0.
\end{align}


\section{Perturbation Theory}
\label{App:perturbation}
\subsection{Normal-ordered expansion of cosinuosidal operators}
To expand the cosinusoidal terms with phase variables $\hat{\varphi_{i}}$ using annihilation ($\hat{a}_i$) and creation ($\hat{a}_i^\dagger$) operators, the terms can be rewritten with displacement operators $\mathcal{D}_i(\alpha) \equiv e^{\alpha \hat{a}_i^\dagger - \alpha^*\hat{a}_i}$ with the corresponding amplitudes $\alpha = \pm i\varphi_i^{zpf}$ as follows:

\begin{align}
\cos{\hat\varphi_i} &= \cos{(\varphi_{i}^{zpf}(\hat{a}_i+\hat{a}_i^\dagger))} \notag\\ &=\frac{1}{2}\left(e^{i\varphi_i^{zpf}(\hat{a}_i+\hat{a}_i^\dagger)} + e^{-i\varphi_i^{zpf}(\hat{a}_i+\hat{a}_i^\dagger)}\right) \notag\\ &=\frac{1}{2}\left(\mathcal{D}(i\varphi_i^{zpf}) + \mathcal{D}(-i\varphi_i^{zpf})\right),
\end{align}

\begin{align}
\sin{\hat\varphi_i} &= \sin{(\varphi_{i}^{zpf}(\hat{a}_i+\hat{a}_i^\dagger))} \notag\\ &=\frac{1}{2i}\left(e^{i\varphi_i^{zpf}(\hat{a}_i+\hat{a}_i^\dagger)} - e^{-i\varphi_i^{zpf}(\hat{a}_i+\hat{a}_i^\dagger)}\right) \notag\\ &=\frac{1}{2i}\left(\mathcal{D}_i(i\varphi_i^{zpf}) -\mathcal{D}_i(-i\varphi_i^{zpf})\right).
\end{align}

\noindent Using the normal-ordered representation obtained from the Baker-Campbell-Hausdorff formula  $\mathcal{D}_i(\alpha) = e^{-\frac{|\alpha|^2}{2}}e^{\alpha \hat{a}_i^{\dagger}}e^{-\alpha^* \hat{a}_i}$ \cite{mccrea1933operational}, we obtain the following normal-ordered expansions:
\begin{align}
    &\cos{\hat\varphi_i} \notag\\
    &=\frac{e^{-\frac{(\varphi_i^{zpf})^2}{2}}}{2}\left(e^{i\varphi_i^{zpf}\hat{a}_i^\dagger}e^{i\varphi_i^{zpf}\hat{a}_i} + e^{-i\varphi_i^{zpf}\hat{a}_i^\dagger}e^{-i\varphi_i^{zpf}\hat{a}_i}\right)\notag\\&=
    e^{-\frac{(\varphi_i^{zpf})^2}{2}}\sum_n{\sum_{k=0}^{2n}{\frac{(-1)^n(\varphi^{zpf}_i)^{2n}}{k!(2n-k)!}(\hat{a}_i^\dagger)^k(\hat{a}_i)^{2n-k}}},\\
    &\sin{\hat\varphi_i} \notag\\
    &=\frac{e^{-\frac{(\varphi_i^{zpf})^2}{2}}}{2i}\left(e^{i\varphi_i^{zpf}\hat{a}_i^\dagger}e^{i\varphi_i^{zpf}\hat{a}_i} - e^{-i\varphi_i^{zpf}\hat{a}_i^\dagger}e^{-i\varphi_i^{zpf}\hat{a}_i}\right)\notag\\ &= e^{-\frac{(\varphi_i^{zpf})^2}{2}}\sum_n{\sum_{k=0}^{2n+1}{\frac{(-1)^n(\varphi^{zpf}_i)^{2n+1}}{k!(2n+1-k)!}(\hat{a}_i^\dagger)^k(\hat{a}_i)^{2n+1-k}}}.
    \label{eq:sinecosine_expansion}
\end{align}
\noindent which are used for calculating matrix elements and perturbative analyses provided in the main text. 

\subsection{Estimating zero-point fluctuations}
Expanding the cosine potential in the bare transmon Hamiltonians $\hat{H}_{i}$ produces the following expression up to the second order:
{
\small
\begin{align}
    \hat{H}_{i} &\approx
    \left(8E_{C,i} (n_{i}^{zpf})^2 - E_{J,i}e^{-\frac{(\varphi_{i}^{zpf})^2}{2}}(\varphi_{i}^{zpf})^2\right)\hat{a}_i^\dagger\hat{a}_i  \notag \\ 
    + &\left(-4E_{C,i} (n_{i}^{zpf})^2 +  \frac{1}{2}E_{J,i}e^{-\frac{(\varphi_{i}^{zpf})^2}{2}}(\varphi_{i}^{zpf})^2\right)(\hat{a}_i^2 + (\hat{a}_i^\dagger)^2).
\end{align}
}

\noindent The condition to eliminate the $\hat{a}_i^2 + (\hat{a}_i^\dagger)^2$ term and the normalization condition $n_i^{zpf}\varphi_{i}^{zpf} = 1/2$ yield the following equation that enables calculating $\varphi_{i}^{zpf}$ numerically:
\begin{align}
(\varphi_i^{zpf})^4 &e^{-\frac{(\varphi_{i}^{zpf})^2}{2}} = \frac{2E_{C,i}}{E_{J,i}}.
\end{align}



\subsection{ZZ interaction rate}

ZZ interaction rate obtained from perturbation theory $\zeta_\text{pert}$ is calculated up to the second-order and single excitation number differences that are considered to provide the leading contributions to the ZZ interaction rate, with the following breakdown:
\begin{equation}
    \zeta_\text{pert} \approx \zeta^{(1)} + \zeta^{(2)}_c + \zeta_\text{odd}^{(2)},
\end{equation}
\noindent where the even-parity interaction is responsible for the first-order contribution $\zeta^{(1)}$ and the second-order contribution $\zeta^{(2)}_c$ due to excitation-number-conserving transitions up to a single-excitation hopping, and the odd-parity interaction due to junction asymmetry accounts for the second-order contribution $\zeta^{(2)}_\text{odd}$ due to single-excitation-number differences.

The $\zeta^{(2)}_\text{odd}$ is calculated as follows:

\begin{align}
    g_{ijkl}&\equiv \bra{ij}\hat{H}_\text{int}|kl\rangle, \ \ \ \omega_{ij} \equiv \bra{ij}\hat{H}_0\ket{ij}, \notag \\ 
    E_{J,C}^\text{sin} &\equiv \Delta E_{J,C}e^{-\frac{(\varphi_{1}^{zpf})^2 + (\varphi_{2}^{zpf})^2}{2}} \sin{\left(\frac{\varphi_{e,1}}{2}\right)},\notag\\
    g_{0100} &=  E_{J,C}^\text{sin}\varphi_2^{zpf}, \ \ \ g_{1000} =  -E_{J,C}^\text{sin}\varphi_1^{zpf}, \notag\\ 
    g_{0201} &=  \sqrt{2}E_{J,C}^\text{sin}\left(\varphi_2^{zpf}- \frac{1}{2}(\varphi_2^{zpf})^3\right), \notag \\ 
    g_{2010} &=  -\sqrt{2}E_{J,C}^\text{sin}\left(\varphi_1^{zpf}- \frac{1}{2}(\varphi_1^{zpf})^3\right), \notag\\ 
    g_{1110} &=  E_{J,C}^\text{sin}\varphi_2^{zpf}\left(1- (\varphi_1^{zpf})^2\right), \notag\\ 
    g_{1101} &=  -E_{J,C}^\text{sin}\varphi_1^{zpf}\left(1- (\varphi_2^{zpf})^2\right), \notag 
\end{align}
\begin{align}
    g_{1211} &=  \sqrt{2}E_{J,C}^\text{sin}\left(1- (\varphi_1^{zpf})^2\right)\left(\varphi_2^{zpf}- \frac{1}{2}(\varphi_2^{zpf})^3\right),\notag\\ 
    g_{2111} &=  -\sqrt{2}E_{J,C}^\text{sin}\left(1- (\varphi_2^{zpf})^2\right)\left(\varphi_1^{zpf}- \frac{1}{2}(\varphi_1^{zpf})^3\right),\notag\\ 
    g_{2001} &=  E_{J,C}^\text{sin}\varphi_2^{zpf}\left(- \frac{\sqrt{2}}{2}(\varphi_1^{zpf})^2\right),\notag\\ 
    g_{0210} &=  -E_{J,C}^\text{sin}\varphi_1^{zpf}\left(- \frac{\sqrt{2}}{2}(\varphi_2^{zpf})^2\right),\notag\\ 
    g_{3011} &=  E_{J,C}^\text{sin}\varphi_2^{zpf}\left(- \frac{\sqrt{6}}{2}(\varphi_1^{zpf})^2 + \frac{\sqrt{6}}{6}(\varphi_1^{zpf})^4\right),\notag\\ 
    g_{0311} &=  -E_{J,C}^\text{sin}\varphi_1^{zpf}\left(- \frac{\sqrt{6}}{2}(\varphi_2^{zpf})^2 + \frac{\sqrt{6}}{6}(\varphi_2^{zpf})^4\right), 
\end{align}

\begin{align}
    &\zeta_\text{odd}^{(2)} = \sum_{ij} (-1)^{i+j} E_{ij, \text{odd}}^{(2)} \notag\\ &= \frac{g_{3011}^2}{\omega_{11} - \omega_{30}} +  \frac{g_{0311}^2}{\omega_{11} - \omega_{03}} - \frac{g_{0210}^2}{\omega_{10} - \omega_{02}} - \frac{g_{2001}^2}{\omega_{01} - \omega_{20}} \notag \\ &+ \frac{g_{2111}^2}{\omega_{11} - \omega_{21}} +  \frac{g_{1211}^2}{\omega_{11} - \omega_{12}} +  \frac{2g_{1110}^2}{\omega_{11} - \omega_{10}} +  \frac{2g_{1101}^2}{\omega_{11} - \omega_{01}}   \notag \\ & - \frac{g_{2010}^2}{\omega_{10} - \omega_{20}} - \frac{g_{0201}^2}{\omega_{01} - \omega_{02}} - \frac{2g_{1000}^2}{\omega_{10} - \omega_{00}} - \frac{2g_{0100}^2}{\omega_{01} - \omega_{00}} 
    \notag \\ &\approx -(E_{J,C}^\text{sin}\varphi_{1}^{zpf}\varphi_{2}^{zpf})^2 \times\left(\frac{(\varphi_{1}^{zpf})^2}{2\omega_1 - \omega_2} + \frac{(\varphi_{2}^{zpf})^2}{2\omega_2 - \omega_1}  \right. \notag\\ & \ \ \ \left. +\frac{4(\varphi_{1}^{zpf})^2}{\omega_1} + \frac{4(\varphi_{2}^{zpf})^2}{\omega_2} + \frac{4\eta_1}{\omega_1^2} + \frac{4\eta_2}{\omega_2^2} \right) \notag \\ 
    &\ \    \sim \mathcal{O}\left(\frac{\Delta E_{J,C}^2(\varphi^{zpf})^6}{m\omega_1 + n\omega_2}\right) + \mathcal{O}\left(\frac{\Delta E_{J,C}^2(\varphi^{zpf})^4\eta}{\omega^2}\right).
\end{align}
\noindent where $\omega_1$ and $\omega_2$ are the bare 0-1 transition frequencies of the two qubits, $(m, n) \in \{(1,0), (0,1), (2,-1), (-1,2)\}$, and the approximations $(\varphi^{zpf})^2 \ll 1$ and $|\eta_1|, |\eta_2| \ll m\omega_1 + n\omega_2$ are used.

\section{Simulation Methods}
\label{App:simulationmethods}

\subsection{Numerical simulation tools}
\texttt{Python} packages \texttt{scQubits} \cite{Groszkowski2021scqubitspython} and \texttt{QuTiP} \cite{johansson2012qutip} are used for numerical circuit quantization, diagonalization, and time-domain simulation. 

\subsection{Computational subspace}
ZZ turn-off point $\Phi_\text{off}$ is found by sweeping $\Phi_{e, 1}$ under the constraint $\Phi_{e,2} = -\Phi_{e,1}/2$. At this point, we perform numerical diagonalization to identify computational eigenstates. The mapping between the dressed eigenstates and the bare eigenstates is found by the hierarchical diagonalization procedure \cite{kerman2020efficientnumericalsimulationcomplex, Chitta2022computeraided}.

\subsection{Gate fidelity}
The coherent error ($1-F_G$) of the CZ gates is estimated using the standard average gate fidelity \cite{pedersen2007fidelity}:
\begin{equation}
    F_G (\hat{U}) = \frac{\text{Tr}(\hat{U}_{proj}^\dagger \hat{U}_{proj}) + |\text{Tr}(\hat{U}_{CZ}^\dagger \hat{U}_{proj})|^2}{20},
\end{equation}
\noindent where $\hat{U}_{proj} \equiv \hat{P}\hat{U}\hat{P}^\dagger$ is the simulated unitary propagator projected onto the computational subspace with the projector $\hat{P}$, and $\hat{U}_{CZ}$ is the ideal unitary for a CZ gate, represented with computational eigenstates. 

For gate infidelities ($1-F$) including incoherent processes, we use the standard gate fidelity involving Kraus operators \cite{pedersen2007fidelity}: 
\begin{equation}
    F (\hat{U}) = \frac{{\text{Tr}(\sum_k \hat{M}_k^\dagger \hat{M}_k) + \sum_k|\text{Tr}(\hat{M}_k)|^2}}{20},
\end{equation}
\noindent where $\hat{M_k} \equiv \hat{P}U_{CZ}\hat{G}_k\hat{P}^\dagger$, and $\hat{G_k}$s are the Kraus operators.

By solving the following optimization problem, we report the coherent error and gate infidelity after eliminating the effect of local phases that can be canceled out by virtual-Z operations \cite{mckay2017virtualz}:
\begin{align}
    \text{min}_{\phi_1, \phi_2} \left(1-F_G (\hat{U}e^{-i\phi_1\hat{Z}_1}e^{-i\phi_2\hat{Z}_2})\right),\\
    \text{min}_{\phi_1, \phi_2} \left(1-F (\hat{U}e^{-i\phi_1\hat{Z}_1}e^{-i\phi_2\hat{Z}_2})\right),
\end{align}
\noindent where $\hat{Z}_1$ and $\hat{Z}_2$ are Pauli-Z operators of the qubits. 


\subsection{Hilbert space truncation}
\begin{figure}[tbp]
\centering
\includegraphics[width = \columnwidth]{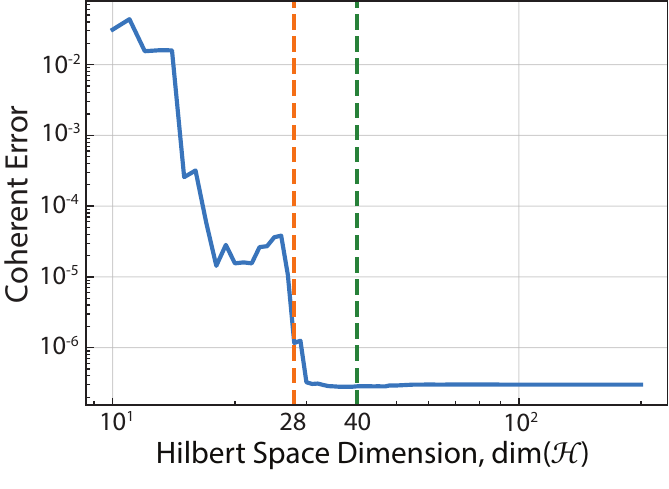}
\caption{\textbf{Effect of Hilbert space truncation.} Coherent error of the simulated 22 ns-long CZ gate as a function of Hilbert space dimension $dim(\mathcal{H})$ (blue solid line). $dim(\mathcal{H}) = 40$ (green dashed line) is used for the time-domain simulations reported in the main text, and $dim(\mathcal{H}) = 28$ (orange dashed line) is used for CZ gate simulation with relaxation process.}
\label{Fig:Hdimtruncation}
\end{figure}

In order to run time-domain simulations within a feasible computation time, the Hilbert space is truncated to include up to the 40 lowest energy eigenstates. To examine the effect of Hilbert space truncation, we sweep the Hilbert space dimension $dim(\mathcal{H})$ and estimate the coherent error of the 22 ns-long CZ gate shown in the main text, as shown in Fig.~\ref{Fig:Hdimtruncation}. For $dim(\mathcal{H}) \geq 40$, the change in the coherent error is less than $2\times10^{-8}$, which is sufficiently small to estimate the coherent errors of $3\times10^{-7}$ reported in the main text. For incoherent simulations in Appendix \ref{App:gatefidelityT1}, we use $dim(\mathcal{H}) = 28$ to reduce computation time, while maintaining the numerical errors from Hilbert space truncation below $10^{-6}$.

\section{Gate fidelity in the presence of relaxation error}
\label{App:gatefidelityT1}

\begin{figure}[tbp]
\centering
\includegraphics[width = \columnwidth]{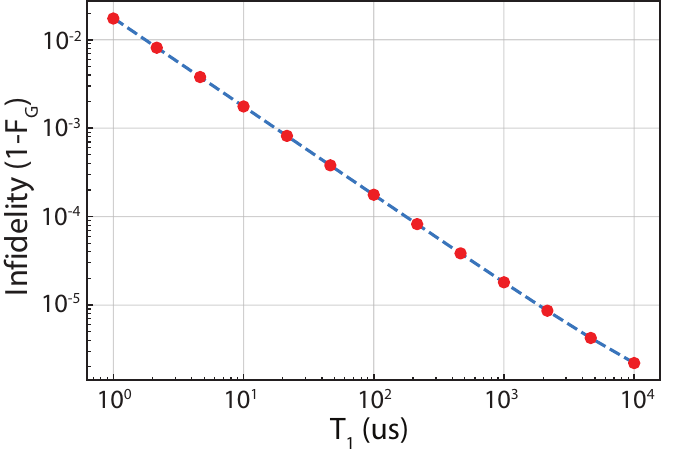}
\caption{\textbf{Gate fidelity in the presence of Markovian relaxation.} Infidelity of the simulated 22 ns-long CZ gate as a function of relaxation time ($T_1$) of the two transmons (red dots). Blue dashed line is a fit to eq.~(\ref{eq:T1errorfit}).} 
\label{Fig:infidloss}
\end{figure}

By simulating the Lindblad master equation, the performance of the CZ gate is examined in the presence of Markovian relaxation, as shown in Fig.~\ref{Fig:infidloss}. We model the relaxation process by adding bare transmon's annihilation operators as jump operators $\sqrt{1/T_1}\hat{a}_i$, where $T_1$ is the bare relaxation time of the transmons. We use the same $T_1$s for both qubits. To construct the bare annihilation operators, we use the hierarchically diagonalization procedure \cite{kerman2020efficientnumericalsimulationcomplex, Chitta2022computeraided}.

The gate infidelity is fitted to a linear model shown in eq.~(\ref{eq:T1errorfit}):

\begin{equation}
    1-F \approx a\frac{T_G}{T_1} + b,
    \label{eq:T1errorfit}
\end{equation}
\noindent recovering $a = 0.799$ and $b = 4.5\times 10^{-7}$, which is consistent with the prediction $a = 0.8$ based on the analysis in Ref.~\cite{pedersen2007fidelity}. 

\section{SQUID couplers with various circuit parameters}
\label{App:circuitparameters}

\subsection{Circuit parameters available for idle operation and robustness against capacitance targeting}
\begin{figure}[tbp]
\centering
\includegraphics[width = \columnwidth]{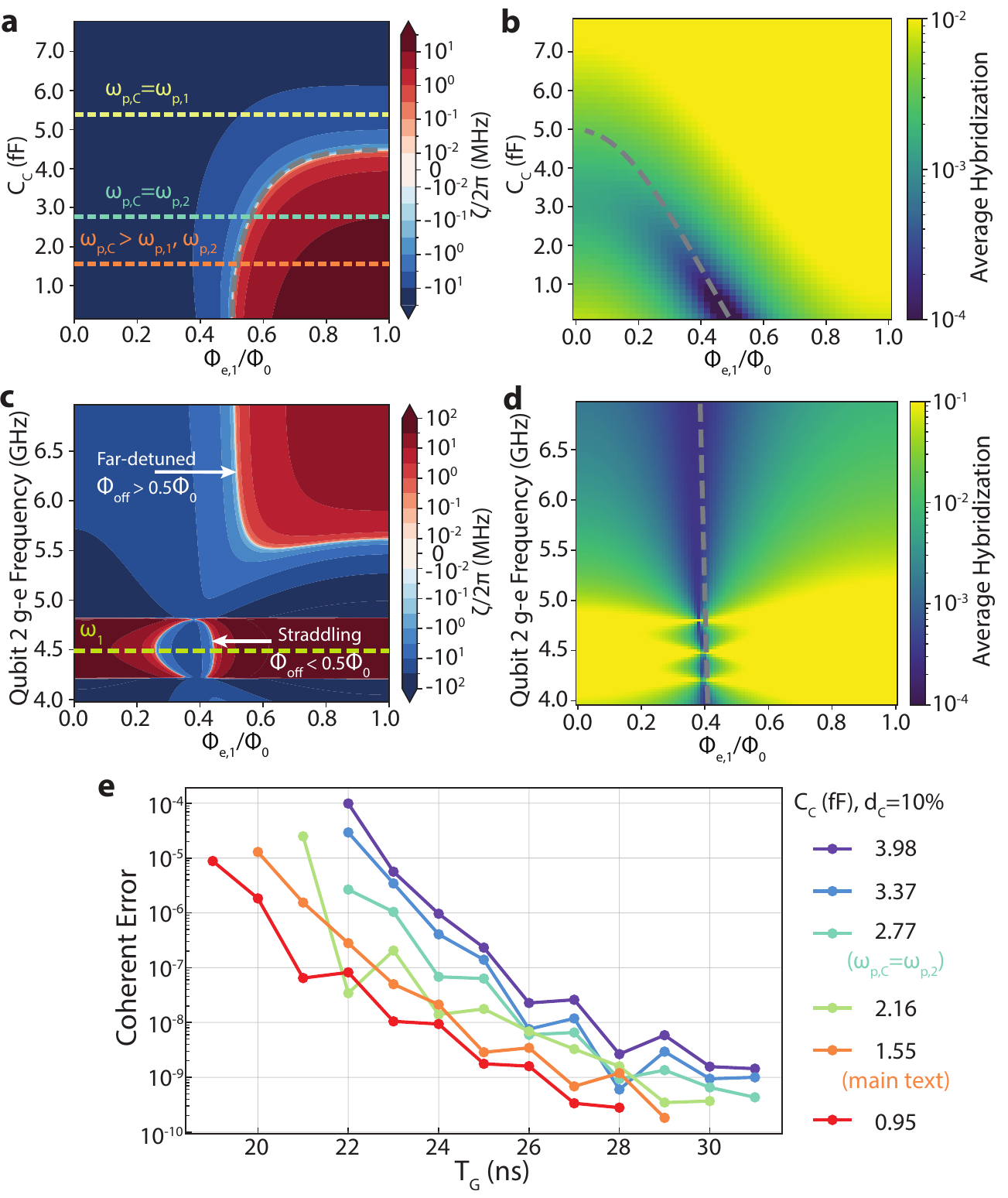}
\caption{\textbf{ZZ interaction rates and hybridization} \textbf{a}, ZZ interaction rate obtained from circuit quantization, as a function of total SQUID coupler capacitance $C_C$ and external flux $\Phi_{e,1}$, while other parameters are the same as the Table. 1 in the main text. Yellow (light blue) dashed line line corresponds to the case where the coupler junctions have the same plasma frequency as transmon 1 (transmon 2). Orange dashed line indicates the capacitance used in the analysis presented in the main text. Gray dashed curve shows the $\Phi_\text{off}$ predicted from perturbative analysis. \textbf{b}, Average hybridization $\sum_{mn}(1-P_{\ket{ij}})/4$ as a function of $C_C$ and $\Phi_{e,1}$. \textbf{c}, ZZ interaction rate as a function of qubit 2 frequency $\omega_{2}$ and $\Phi_{e,1}$, with other parameters are the same as the Table. 1 in the main text. $\omega_2$ is varied by changing $E_{J,2}$. Yellow dashed line represents qubit 1 frequency $\omega_1$. \textbf{d}, Average hybridization as a function of $\omega_{2}$ and $\Phi_{e,1}$. In \textbf{b} and \textbf{d}, Gray dashed lines indicate the biases $\Phi_{e,1}$ where the effective linear coupling is canceled $g_\text{eff}(\Phi_{e,1}) = 0$, predicted from the perturbation theory. \textbf{e}, Coherent errors of the simulated CZ gates as functions of gate duration $T_G$, for different $C_C$.} 
\label{Fig:circuitparams}
\end{figure}

The existence of at least one idle operating point $\Phi_\text{off}$ at which the ZZ interaction rate is turned off $\zeta(\Phi_\text{off}) = 0$ is crucial. This condition can be found approximately by investigating the following expression obtained from the second-order perturbation theory:
{\small
\begin{align}
    \zeta_\text{pert}(\varphi_{e,1}) &= -\Sigma E_{J,C}'\cos{\left(\frac{\varphi_{e,1}}{2}\right)} (\varphi_1^{zpf})^2(\varphi_2^{zpf})^2 + \frac{4g_\text{eff}^2\eta }{\Delta^2 - \eta^2} , \\
    g_\text{eff} &= -\Sigma E_{J,C}'\cos{\left(\frac{\varphi_{e,1}}{2}\right)}\varphi_1^{zpf}\varphi_2^{zpf} + gn_1^{zpf}n_2^{zpf}. \label{eq:perturbationzzinteraction}
\end{align}}

\noindent When the two qubits are far-detuned, $|\Delta| \gg |\eta|$, the sign of the second term in $\zeta_\text{pert}$ is negative, as transmon anharmonicities are negative. Thus, the ZZ interaction rate can be turned off only at $\varphi_{e,1} > \pi$ at which the first term becomes positive. As $\zeta_\text{pert}(0) > 0$ is always satisfied, the following condition guarantees the existence of an idling external flux due to the intermediate value theorem:

\begin{align}
     0 \leq \zeta_\text{pert}(2\pi) &= \Sigma E_{J,C}' (\varphi_1^{zpf})^2(\varphi_2^{zpf})^2 \notag\\ &+ \frac{4(\Sigma E_{J,C}'\varphi_1^{zpf}\varphi_2^{zpf} + gn_1^{zpf}n_2^{zpf} )^2\eta }{\Delta^2 - \eta^2}.
\end{align}

\noindent As the first term is always positive, the second term originating from the hybridization, which is negative, should be suppressed. This is achieved by reducing the charge coupling rate $g$ and increasing the absolute value of qubit-qubit detuning $|\Delta|$. 

In contrast, when the two transmons are in the straddling regime $\omega_{02}, \omega_{20} < \omega_{11}$ or $|\Delta| < |\eta|$, the second term in eq.~(\ref{eq:perturbationzzinteraction}) becomes positive and the idle operating point would be located at $\varphi_{e,1} < \pi$, if it exists.

Fig.~\ref{Fig:circuitparams}a shows the ZZ interaction rates as functions of external flux calculated using circuit quantization and numerical diagonalization, as $C_C$ is swept. We find that at sufficiently small $C_C$s, the idle operating point exists. Comparison of the plasma frequency $\omega_{p,C}$ of the coupler junctions to the plasma frequencies of the qubit junctions' $\omega_{p,1}$ and $\omega_{p,2}$ shows that an idle operating point can exist without requiring extreme oxidation conditions to lower $\omega_{p, C}$ further. The existence of an idle operating point at $\omega_{p,C} = \omega_{p,2}$ and its robustness against $C_C$ implies that the same junction oxidation condition used in one of the qubits can be used for junctions in SQUID couplers. The average hybridization can be suppressed to or below 1\% for a similar range of $C_C$s that provides an idle bias point, as found in Fig.~\ref{Fig:circuitparams}b. In addition, such junction processes with thick oxide barriers ease junction area targeting due to their smaller tunneling energy per area. 

Fig.~\ref{Fig:circuitparams}c shows the ZZ interaction rates as the qubit 2's g-e transition frequency is swept, by varying $E_{J,2}$. Qubit 1's g-e frequency is approximately at 4.49 GHz. As expected, $\Phi_\text{off} > 0.5\Phi_{0}$ $(\varphi_{e,1} > \pi)$ exists for sufficiently large detuning, and $\Phi_\text{off} < 0.5\Phi_{0}$ $(\varphi_{e,1} < \pi)$ in the straddling regime.

To examine the tolerance of SQUID coupler operation against circuit parameter targeting inaccuracy and to explore available junction plasma frequencies for the junctions in the SQUID coupler, we estimate the coherent errors of the CZ gates for different $C_C$ for the 10 shortest gate durations, as shown in Fig.~\ref{Fig:circuitparams}e. We assume an asymmetry $d_C = \Delta E_{J,C}/{\Sigma E_{J,C}} = (C_{C,1}-C_{C,2})/C_C = 10$ \% for the two junctions in the SQUID couplers, in which we correlate the asymmetry in the junction energies and the capacitance. We find that all $C_C$s in the wide range of [0.95, 3.98] fF provide fast and high-fidelity CZ gates with small adiabaticity overheads, showing the robustness of the gate scheme against coupler capacitance targeting. Note that the difference in the minimum CZ gate time is due to the change in the hybridization-induced ZZ interaction rate $\zeta^{(2)}$ originating from charge coupling.

\subsection{Junction coupler}

\begin{figure}[tbp]
\centering
\includegraphics[width = \columnwidth]{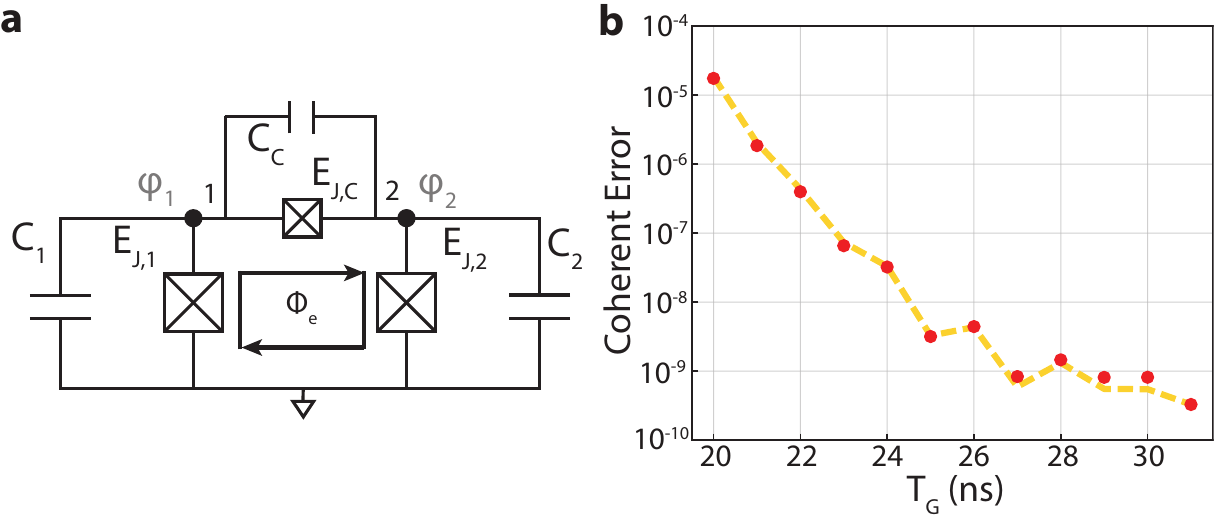}
\caption{\textbf{CZ gate with junction coupler.} \textbf{a}, Circuit diagram of the junction coupler. \textbf{b}, Coherent error of the CZ gate as a function of gate duration $T_G$, implemented by the junction coupler. Dashed yellow line represents the contribution from the non-adiabatic state transitions.} 
\label{Fig:junctioncoupler}
\end{figure}

\begin{figure}[tbp]
\centering
\includegraphics[width = \columnwidth]{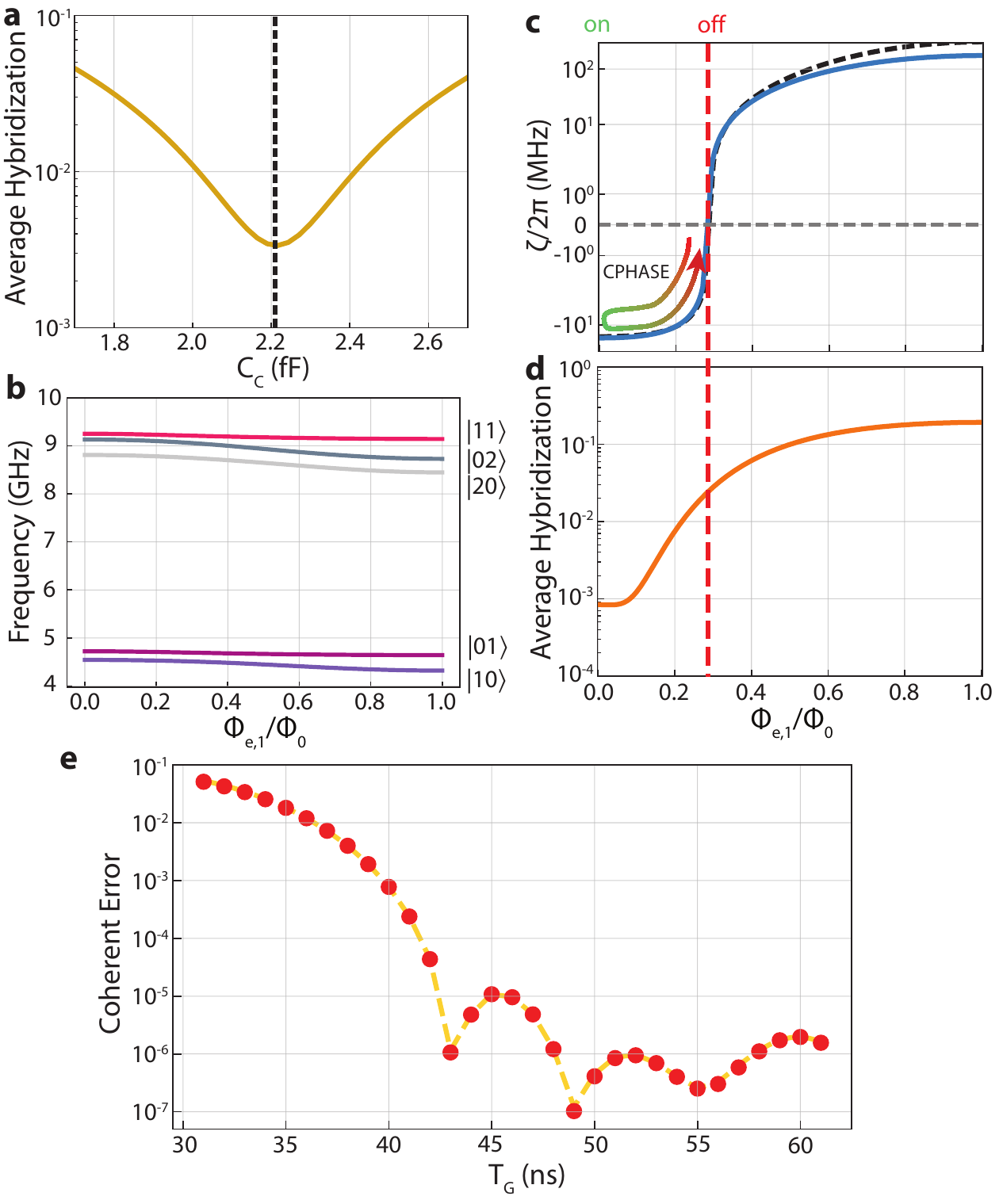}
\caption{\textbf{Two transmons in straddling regime coupled via a SQUID coupler.} \textbf{a}, Average hybridization as a function of total coupler junction capacitance $C_C$, at $\Phi_{e,1} = 0$. Dashed black line denotes the $C_C$ that provides the minimum hybridization. \textbf{b}, Eigenfrequencies of the five lowest eigenstates as functions of external flux $\Phi_{e,1}$. \textbf{c, d} ZZ interaction rate and average hybridization as functions of $\Phi_{e,1}$. Dashed red line represents the ZZ interaction turn-off point $\Phi_\text{off}$. \textbf{e}, Coherent error of the CZ gate implemented with the straddling regime parameter. } 
\label{Fig:straddling}
\end{figure}

The limit in which the asymmetry is 100 percent or -100 percent corresponds to the case in which one of the junctions in the SQUID coupler is eliminated, as shown in Fig.~\ref{Fig:junctioncoupler}a. The interaction Hamiltonian is thus provided as follows:
\begin{align}
    \hat{H}_\text{int}  & \approx 
     -E_{J,C}\cos{\left(\hat{\varphi}_2 - \hat{\varphi}_1 + \varphi_{e}\right)}
     +g\hat{n}_1\hat{n}_2 \notag \\ 
     &=
     -E_{J,C}\cos{(\varphi_{e})}{\left(\hat{\varphi}_2 - \hat{\varphi}_1\right)} \notag\\ &+ E_{J,C}\sin{(\varphi_{e})}{\left(\hat{\varphi}_2 - \hat{\varphi}_1\right)}
     +g\hat{n}_1\hat{n}_2, 
    \label{eq:hamiltonian_junction_coupler}
\end{align}
\noindent where $\varphi_e \equiv 2\pi\Phi_{e}/\Phi_{0}$. The SQUID coupler interaction Hamiltonian is mapped to the junction coupler interaction Hamiltonian under $\Sigma E_{J,C}, \Delta E_{J,C} \xrightarrow{} E_{J,C}$ and $\Phi_{e,1} + \Phi_{e,2} \xrightarrow{} \Phi_{e}$. Defining $\Phi_{e,1} \equiv -2\Phi_{e}$, we can compare the junction coupler with the SQUID coupler in a consistent way.

The junction coupler is an extreme example that shows the robustness of the SQUID coupler against the junction asymmetry. The junction coupler still yields an idle external flux and allows the implementation of a high-fidelity adiabatic CZ gate as efficient as the symmetric SQUID coupler, whose coherent error is presented in Fig.~\ref{Fig:junctioncoupler}b. However, they induce longitudinal interactions that act as the crosstalk channel when chained with more couplers and qubits, as discussed in the main text.

\subsection{CZ gate in straddling regime}

The SQUID coupler can be used for implementing a CZ gate between two transmons in a stradling regime. In this example, the transmon 2 frequency $\omega_2$ is brought close to $\omega_{1}$ to satisfy the straddling condition $\omega_{\ket{20}}, \omega_{\ket{02}}< \omega_{\ket{11}}$, as shown in Fig.~\ref{Fig:straddling}b. In this parameter regime, it is crucial to suppress linear coupling, due to their relatively large contribution to the ZZ interaction rate. We choose $C_C$ that minimizes the average hybridization at $\Phi_{e} = 0$, as found in Fig.~\ref{Fig:straddling}a. 

As discussed earlier, the idle external flux $\Phi_\text{off}$ is less than $0.5\Phi_{0}$ in the straddling regime, as shown in Fig.~\ref{Fig:straddling}c. Similarly to the pair of transmons that are far-detuned, a CZ gate can be implemented by dynamically tuning the external flux. We find that a 43 ns-long CZ gate with coherent error of $10^{-6}$ can be implemented as illustrated in Fig.~\ref{Fig:straddling}e, showing the utility of the SQUID coupler in the straddling regime.


\section{Calculation of losses from interfacial dielectric layers}

\label{App:interfacialloss}

\begin{figure}[tbp]
\centering
\includegraphics[width = \columnwidth]{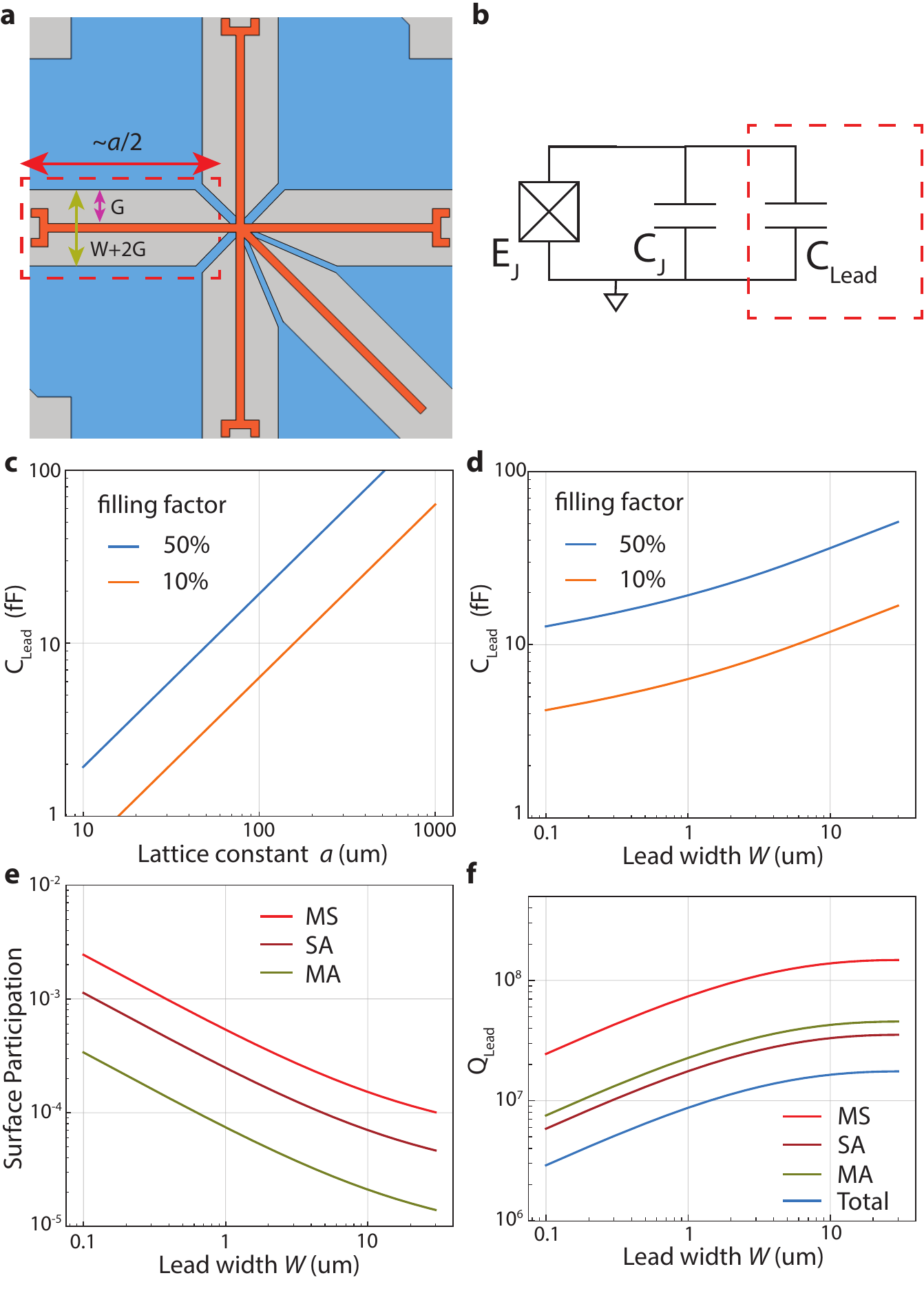}
\caption{\textbf{Calculation of the losses from the interfacial dielectric layers.} \textbf{a}, A single mergemon modeled as a cross-shaped metalization (orange), where the mergemon junction is located at the center and the other parts of the metalization is considered ``leads". Ground metalization (blue) is connected to the low voltage side metalization of the mergemon at the center (hidden). \textbf{b}, Circuit schematic of the mergemon. $C_J$ denotes the contribution from the junction, and $C_{Lead}$ indicates the contribution from the leads. \textbf{c}, Analytically calculated $C_{Lead}$ as a function of the lattice constant. A gap to ground $G$ of 8 um and a lead width $W$ of 1 um are used. \textbf{d}, Analytically calculated $C_{Lead}$ as a function of the lead width. A gap to ground $G$ of 8 um and a lattice constant of 100 um are used. \textbf{e}, Analytically calculated surface participation as a function of the lead width using the formula provided in Ref.~\cite{murray2018analytical}. Trenching factors of 1/4, 1/3, and 4 are multiplied to MS, SA, and MA interfaces, respectively. \textbf{f}, Analytically calculated $Q_{Lead}$ and the contributions to $Q_{Lead}$ from each interfacial layers as functions of the lead width.}
\label{fig:interfacial_loss}
\end{figure}

As the SQUID coupler does not require capacitive coupling from the planar capacitance, the capacitance to ground from the lead for galvanic connection can be suppressed. This allows us to use narrower metalizations and gaps to the ground while keeping the losses from contaminated interfacial dielectric layers small. The interfacial loss due to the geometry presented in Fig.~\ref{fig:interfacial_loss}a is modeled by approximating the leads to coplanar waveguides (CPWs) of a uniform geometry, with the lead width $W$ and the gap to ground $G$ being the CPW parameters.

The capacitance of a single mergemon to ground can be broken down into junction capacitance $C_J$ and lead capacitance $C_{Lead}$, as illustrated in the circuit diagram provided in Fig.~\ref{fig:interfacial_loss}b. Thus, the contribution of the leads to the quality factor $Q_{Lead}$ can be modeled as the following:

\begin{align}
    \frac{1}{Q_{Lead}} &= \frac{U_{Lead}}{U_{tot}}\sum_{i \in \{MS, SA, MA\}}\frac{U_{i}}{U_{Lead}}\tan{\delta_i} \notag\\&\approx \frac{C_{Lead}}{C_\Sigma}\sum_{i \in \{MS, SA, MA\}}p_{i, CPW}\tan(\delta_{i}).
    \label{eq:qfactorsurface}
\end{align}

\noindent where $U_{Lead}$ and $U_{tot}$ are the energy stored in leads and the total energy, respectively, and $U_{i}$ is the energy stored in the interfacial layer $i$, $p_{i, CPW}$ is the surface participation for interfacial layer $i$ calculated by approximating the lead geometry as a CPW assuming that the entire energy is stored in the leads (i.e., the denominator is $U_{Lead}$), $\tan{\delta_i}$ is the corresponding loss tangent, and $C_\Sigma = C_J + C_{Lead}$ is the total capacitance of the mergemon. We consider the three interfacial layers: metal-to-substrate (MS), substrate-to-air (SA), and metal-to-air (MA).

\begin{figure*}[tbp]
\centering
\includegraphics[width = \textwidth]{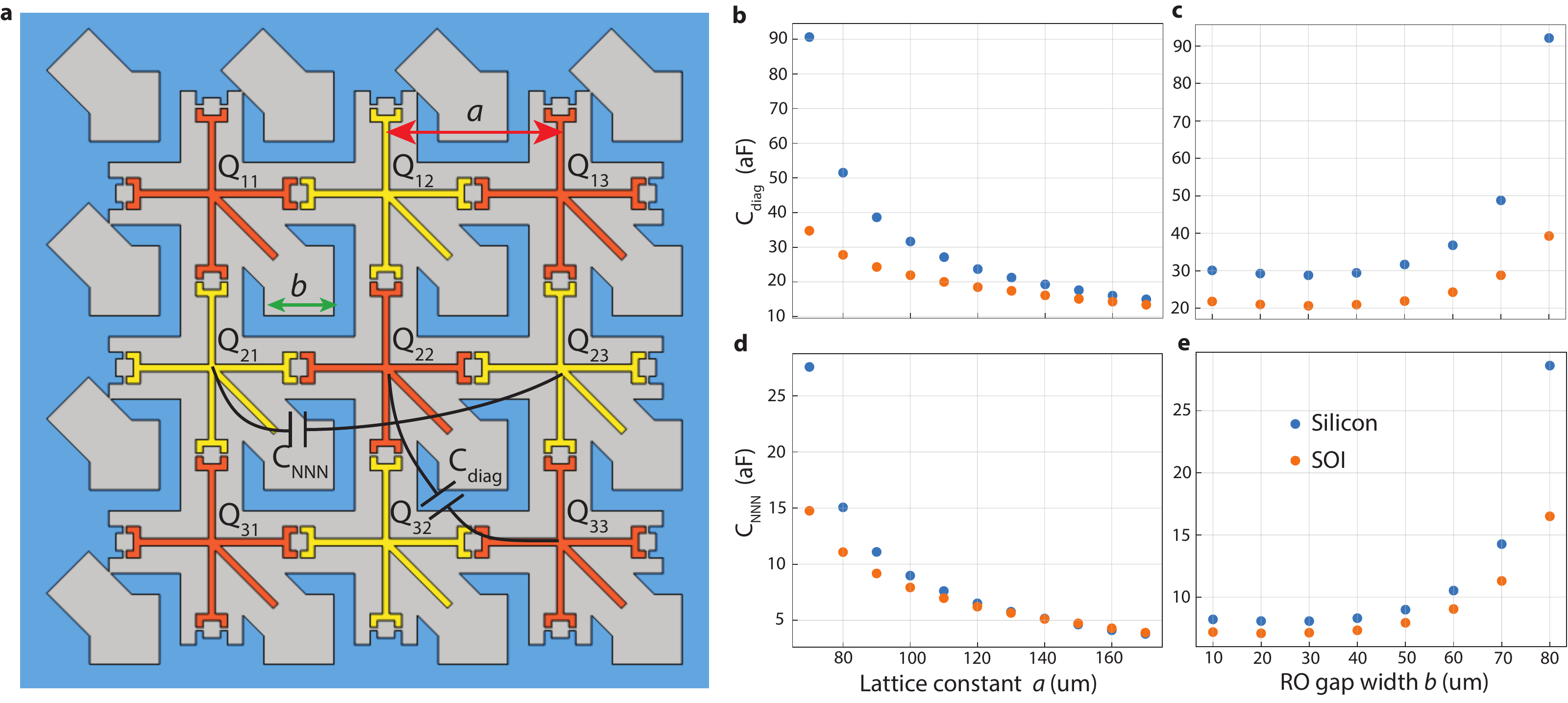}
\caption{\textbf{Parasitic capacitance estimation.} \textbf{a}, Geometry for electrostatics simulation. Lattice constant ($a$), RO gap width ($b$), parasitic capacitances $C_{diag}$ and $C_{NNN}$ are denoted. \textbf{b, d}, $C_{diag}$ and $C_{NNN}$ as functions of lattice constant, for RO gap width = 50 um.  \textbf{c, e}, $C_{diag}$ and $C_{NNN}$ as functions of RO gap width, for lattice constant = 100 um. Simulations are conducted with two different substrate material settings: silicon and SOI.}
\label{fig:parasiticcap_static_sim}
\end{figure*}

We provide the analytically calculated $C_{Lead}$ in Fig.~\ref{fig:interfacial_loss}b and c as a function of two parameters: the lattice constant ($a$) and the lead width $W$, following the formula provided in Ref.~\cite{simons2004coplanar}, for silicon substrate with $\epsilon_{si} = 11.47$ and gap to ground $G = 8$ um. In this model, $C_{Lead}$ is directly proportional to the total length of the leads, which we assume to be 2.5$\times$lattice constant, considering the four connections to the nearest-neighbor qubits and the connection to the readout resonator, as illustrated in Fig.~\ref{fig:interfacial_loss}a. To reduce $C_{Lead}$, we consider the trenching of the substrate and assume a filling factor of 10\%. The following analysis assumes a lattice constant of 100 um.

Surface participation as a function of lead width is calculated by approximating the lead geometry as a CPW and using the analytical model provided in Refs.~\cite{murray2018analytical}, as shown in Fig.~\ref{fig:interfacial_loss}. We assume a uniform thickness of 2 nm for the contaminated dielectric for all interfacial layers. MS, SA, and MA relative permittivities of 11.4, 10, and 4, respectively, are used to consistently cite the loss tangents provided in Ref.~\cite{melville2020comparisondielectric}. To account for the effect of substrate trenching that reduces MS and SA participations and increases MA participation \cite{calusine2018improving, murray2020analyticaltrenched}, the MS, SA, and MA participations are multiplied by 1/4, 1/3, and 4, respectively.

Finally, the contributions to the quality factor are calculated using eq.~(\ref{eq:qfactorsurface}), assuming $C_\Sigma$ of 80 fF, shown in Fig.~\ref{fig:interfacial_loss}. The MS and SA loss tangents are obtained from \cite{melville2020comparisondielectric}. Considering using Tantalum films for the leads, which are known to provide one of the cleanest MA surfaces \cite{place2021newmaterial, shi2022tantalum, crowley2023tantalum, marcaud2025lowlosstantalum, vanschijndel2025cryogenictantalum}, we use the MA loss tangent provided in Ref.~\cite{crowley2023tantalum} with consistent scaling considering the interfacial layer thickness and the relative permittivity as suggested by Ref.~\cite{calusine2018improving}. As expected, the quality factor improves as the lead widens due to the decrease in surface participation, which stagnates at $G \ll W$ as the increase in $C_{Lead}$ catches up. For a lead width of 1 um, we find $Q_{Lead} \approx 8.7 \times 10^{6}$, whose contribution to the relaxation time is $T_1^{Lead} = 1/(\omega Q_{Lead})\approx 309$ us for a $\omega/2\pi = 4.5$ GHz qubit.

\section{Parasitic Capacitance}
\label{App:parasiticcaps}

The parasitic capacitances in the proposed tiling scheme is estimated by finite-element method (FEM), using the COMSOL electrostatic simulation module. We model the metalizations in the mergemon processor as shown in Fig.~\ref{fig:parasiticcap_static_sim}a. Each colored cross (orange, yellow) with a label $Q_{ij}$ represents the high voltage side metalization of the mergemon at $(i,j)$ location, including the elongated leads for galvanic connection to the SQUID couplers. We assume lead thicknesses of 1um. The placeholders for SQUIDs are modeled as capacitors between nearest-neighbor qubits, with footprints of 5 um $\times$ 5 um and gaps between high voltage side leads of 1um. The distance between the leads and the ground metalization (blue) is 8um. We consider two different substrates: 1) a 500 um-thick silicon substrate with relative permittivity of $\epsilon_r = 11.47$, and 2) silicon-on-insulator (SOI) substrate with a 220 nm-thick suspended silicon layer, a 3 um-thick vacuum gap layer, and a 725 um-thick silicon handle layer \cite{keller2017soiqubit}. For both substrates, the layer above the metalizations is set to vacuum. 

In this model, the elongated leads for realizing the galvanic connection are expected to determine the bulk of the capacitance. We identify two crucial dimensions that contribute to parasitic capacitances. First, the lattice constant ($a$) determines the proximity among the leads and the perimeter of the leads. Second, the width of the gapped regions to locate the lumpled element readout resonators (RO gap width, $b$) determines the amount of ground metalization around each lattice site for a given lattice constant, which screens the eletric field. Note that the leads extended from the low voltage metalization of the mergemons are omitted to reduce computation overhead, omission of which would likely lead to overestimation of the parasitic capacitances.

Fig.~\ref{fig:parasiticcap_static_sim}b-e show the estimated parasitic capacitances as functions of the aforementioned parameters. We plot the two largest parasitic capacitances: the diagonal parasitic capacitance $C_{diag}$ between the lattice sites $(i,j)$ and $(i+1, j+1)$ and the next-nearest-neighbor parasitic capacitance $C_{NNN}$ between lattice sites $(i,j)$ and $(i+2, j)$ or between lattice sites $(i,j)$ and $(i, j+2)$, as illustrated in Fig.~\ref{fig:parasiticcap_static_sim}a.

We find that increasing the lattice constant reduces parasitic capacitances, for the same RO gap width. This indicates that the reduction in the parasitic capacitances, due to the increase in the ground metalization area, is faster than the increase in the parasitic capacitance originating from the longer lead lengths. The increase in the RO gap width significantly strengthens the parasitic capacitances by reducing the ground metalization area. For the combination of parameters that yields a sufficient amount of ground metalization (large $a$, small $b$), the parasitic capacitances of the two different substrates become closer in value. This implies that the bulk of the parasitic capacitances are mediated by the stray capacitances in the vacuum layer above the metalizations in such parameter regimes. We find that for the lattice constant of 100 um and the RO gap widths of 50 um, parasitic capacitances are approximately 30 aF or lower for both substrate materials. As mentioned in the main text, simulations including substrate trenching and galvanic connection elements to an opposing chip are expected to estimate smaller stray capacitances by reducing substrate filling factors and adding more ground metalizations that screen electric fields.

\section{Tiling Transmons with Shunt Capacitors and Dual-rail Transmons}

\label{App:tiling}
\begin{figure*}[tbp]
\centering
\includegraphics[width = \textwidth]{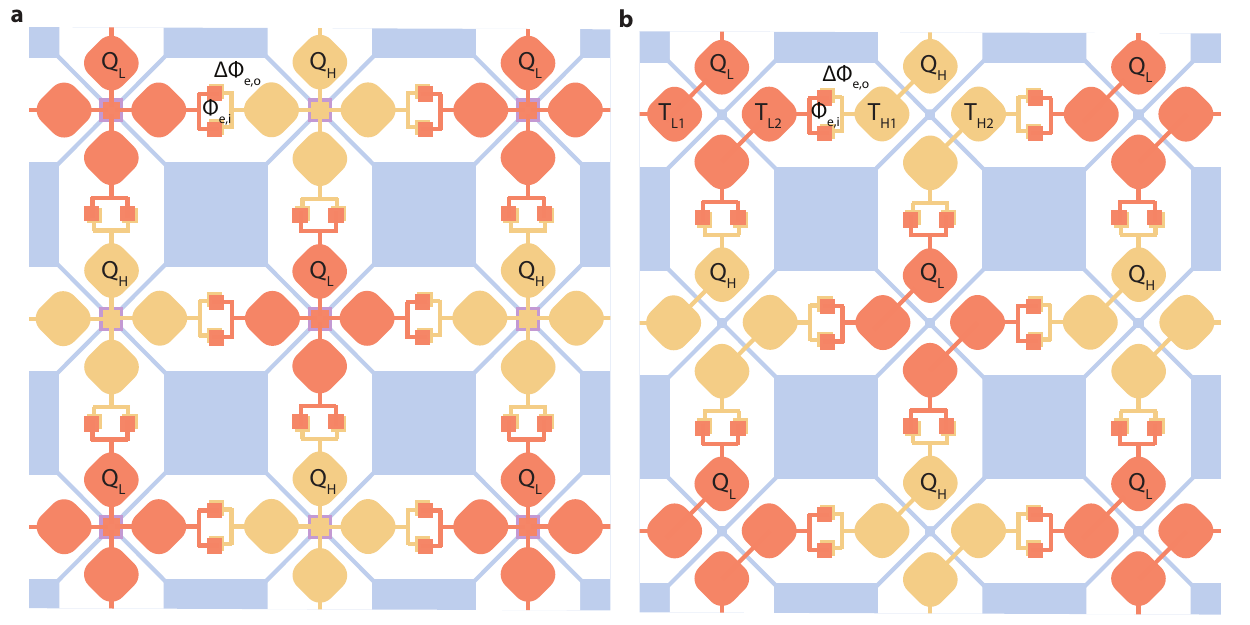}
\caption{\textbf{Tiling transmons with shunt capacitors and dual-rail transmons on a square lattice.} \textbf{a}, Transmons tiled on a square lattice. High voltage metalization (yellow, orange) provides the shunt capacitor needed for transmon. Josephson junctions are located at the center of each unit cell, illustrated as the overlap between the high voltage metalization and the low voltage metalization (purple) that are each galvanically connected to the four extended leads. \textbf{b}, Dual-rail transmons tiled on a square lattice. Each unit cell consists of two transmons, which are coupled capacitively. The overlap between the high voltage leads (yellow, orange) and the ground metalization (blue) represents a single Josephson junction needed for implementing a fixed-frequency transmon or two Josephson junctions forming a SQUID loop needed for implementing a frequency-tunable transmon.}
\label{fig:regular_dualrail_tiling}
\end{figure*}

The tiling strategy for conventional transmons with planar shunt capacitance can be obtained simply by widening the leads and gaps to ground starting from the mergemon tiling scheme, as illustrated in Fig.~\ref{fig:regular_dualrail_tiling}a. The presence of shunt capacitors requires the outer SQUID loops to be enlarged, which poses a trade-off between dephasing due to flux noise in the outer SQUID loops and the dielectric loss from the contaminated interfacial layers on capacitor pads. The bulky capacitor pads may also mediate larger stray capacitance, which can lead to more spectator ZZ interactions.

In contrast, the use of dual rail transmons suppresses the effect of flux noise, due to the protection against frequency noise provided by their operation at artificial sweet-spots \cite{kubica2023erasure, levine2024dualrail}. We provide a tiling example with dual-rail transmons in Fig.~\ref{fig:regular_dualrail_tiling}b, where each unit cell consists of two capacitively coupled transmons. 

\vfill
\bibliography{bib}

\end{document}